\documentclass[twocolumn]{aastex63}
\usepackage{graphics}
\usepackage{color}
\usepackage{topcapt}
\usepackage{amsmath,amstext}
\usepackage{empheq} 
\usepackage{hyperref}
\newcommand{\ltsimeq}{\la}
\newcommand{\gtsimeq}{\ga}

\newcommand{\msun}{M$_{\odot}$}
\newcommand{\mstar}{$M_*$}

\newcommand{\zsun}{Z$_{\odot}$}

\newcommand{\tten}{$\tau_{10}$}
\newcommand{\ttwentyfive}{$\tau_{25}$}
\newcommand{\tfifty}{$\tau_{50}$}
\newcommand{\teighty}{$\tau_{80}$}
\newcommand{\tninety}{$\tau_{90}$}
\newcommand{\hi}{H{\sc i}}
\newcommand{\hii}{H{\sc ii}}

\newcommand{\lp}{Leo~P}

\definecolor{amaranth}{rgb}{0.9, 0.17, 0.31}

\shortauthors{McQuinn et al.}\shorttitle{The Ancient SFH of \lp}

\begin{document}
\title{The Ancient Star Formation History of the Extremely Low-Mass Galaxy Leo~P: An Emerging Trend of a Post-Reionization Pause in Star Formation}

\author[0000-0001-5538-2614]{Kristen B.~W. McQuinn}
\affiliation{Department of Physics and Astronomy, Rutgers, The State University of New Jersey, 136 Frelinghuysen Rd, Piscataway, NJ 08854, USA}
\affiliation{Space Telescope Science Institute, 3700 San Martin Drive, Baltimore, MD 21218, USA}
\email{kmcquinn@stsci.edu}

\author[0000-0002-8092-2077]{Max J.~B. Newman}
\affiliation{Department of Physics and Astronomy, Rutgers, The State University of New Jersey, 136 Frelinghuysen Rd, Piscataway, NJ 08854, USA}

\author[0000-0003-0605-8732]{Evan D. Skillman}
\affiliation{University of Minnesota, Minnesota Institute for Astrophysics, School of Physics and Astronomy, 116 Church Street, S.E., Minneapolis,
MN 55455, USA}

\author[0000-0003-4122-7749]{O. Grace Telford}
\altaffiliation{Carnegie-Princeton Fellow}
\affiliation{Department of Astrophysical Sciences, Princeton University, 4 Ivy Lane, Princeton, NJ 08544, USA}
\affiliation{The Observatories of the Carnegie Institution for Science, 813 Santa Barbara Street, Pasadena, CA 91101, USA}
\affiliation{Department of Physics and Astronomy, Rutgers, The State University of New Jersey, 136 Frelinghuysen Rd, Piscataway, NJ 08854, USA}

\author[0000-0002-0372-3736]{Alyson Brooks}
\affiliation{Department of Physics and Astronomy, Rutgers, The State University of New Jersey, 136 Frelinghuysen Rd, Piscataway, NJ 08854, USA}
 \affiliation{Center for Computational Astrophysics, Flatiron Institute, 162 Fifth Ave, New York, NY 10010, USA}

\author[0000-0002-9798-5111]{Elizabeth A. K. Adams}
\affiliation{ASTRON, The Netherlands Institute for Radio Astronomy, Oude Hoogeveensedijk 4, 7991 PD, Dwingeloo, The Netherlands}
\affiliation{Kapteyn Astronomical Institute, University of Groningen, Postbus 800, 9700 AV Groningen, The Netherlands}

\author[0000-0002-4153-053X]{Danielle A. Berg}
\affiliation{Department of Astronomy, The University of Texas at Austin, 2515 Speedway, Stop C1400, Austin, TX 78712, USA}

\author[0000-0003-4850-9589]{Martha L. Boyer}
\affiliation{Space Telescope Science Institute, 3700 San Martin Drive, Baltimore, MD 21218, USA}

\author[0000-0002-1821-7019]{John M. Cannon}
\affiliation{Department of Physics and Astronomy, Macalester College, 1600 Grand Avenue, Saint Paul, MN 55105, USA}

\author[0000-0001-8416-4093]{Andrew E. Dolphin}
\affiliation{Raytheon, 1151 E. Hermans Road, Tucson, AZ 85756, USA}
\affiliation{University of Arizona, Steward Observatory, 933 North Cherry Avenue, Tucson, AZ 85721, USA}

\author[0000-0003-4464-4505]{Anthony J. Pahl}
\altaffiliation{Carnegie Fellow}
\affiliation{The Observatories of the Carnegie Institution for Science, 813 Santa Barbara Street, Pasadena, CA 91101, USA}

\author[0000-0001-8283-4591]{Katherine L. Rhode}
\affiliation{Department of Astronomy, Indiana University, 727 East Third Street, Bloomington, IN 47405, USA} 

\author[0000-0001-8483-603X]{John J. Salzer}
\affiliation{Department of Astronomy, Indiana University, 727 East Third Street, Bloomington, IN 47405, USA} 

\author[0000-0002-2970-7435]{Roger E. Cohen}
\affiliation{Department of Physics and Astronomy, Rutgers, The State University of New Jersey, 136 Frelinghuysen Rd, Piscataway, NJ 08854, USA}

\author[0000-0002-8937-3844]{Steve R. Goldman}
\affiliation{Space Telescope Science Institute, 3700 San Martin Drive, Baltimore, MD 21218, USA}

\begin{abstract}
Isolated, low-mass galaxies provide the opportunity to assess the impact of reionization on their star formation histories (SFHs) without the ambiguity of environmental processes associated with massive host galaxies. There are very few isolated, low-mass galaxies that are close enough to determine their SFHs from resolved star photometry reaching below the oldest main sequence turnoff. JWST has increased the volume for which this is possible, and here we report on JWST observations of the low-mass, isolated galaxy Leo~P.  From NIRCam imaging in F090W, F150W, and F277W, we derive a SFH which shows early star formation followed by a pause subsequent to the epoch of reionization which is then later followed by a re-ignition of star formation. This is very similar to the SFHs from previous studies of other dwarf galaxies in the ``transition zone'' between quenched very low-mass galaxies and the more massive galaxies which show no evidence of the impact of reionization on their SFHs; this pattern is rarely produced in simulations of SFHs. The lifetime SFH reveals that Leo~P's stellar mass at the epoch of reionization was in the range that is normally associated with being totally quenched. The extended pause in star formation from $z\sim5-1$ has important implications for the contribution of low-mass galaxies to the UV photon budget at intermediate redshifts. We also demonstrate that, due to higher sensitivity and angular resolution, observing in two NIRCam short wavelength filters is superior to observing in a combination of a short and a long wavelength filter.
\end{abstract} 

\section{Introduction}\label{sec:intro}
\subsection{Local, Isolated, Low-Mass Galaxies: A Precious Sample}
In the hierarchical model of structure formation, galaxies start small, grow through accretion and mergers, and transform into the rich diversity of galaxies seen at the present-day. Yet, low-mass (\mstar $<10^9$ \msun) galaxies remain the most common type of galaxy by far. Indeed, galaxies with \mstar $<10^7$ \msun\ are estimated to make up 75\% of galaxies in the nearby universe \citep{Martin2019}. Not only are low-mass galaxies the most prevalent galactic structures in the universe, they are also extremely sensitive to internal and external perturbations and therefore are excellent probes of myriad physical processes \citep[e.g.,][]{Bullock2017}. Of particular interest is the degree to which these small `seed' galaxies are shaped by the epoch of reionization, how they subsequently evolve, and the role that environment plays in their growth.

One of the main parameters critical in understanding how various factors govern and/or influence the growth of galaxies is their mass. From a theoretical perspective, galaxies below a dark matter halo mass corresponding to a virial temperature of $\sim10^4$ K (i.e., M$_{\rm halo}$ $\sim10^8$ \msun) are expected to be quenched by reionization regardless of environment, likely with an assist by stellar feedback. In the shallow potential wells of such systems, the galaxies are unable to self-shield and the ultraviolet (UV) photons from reionization and feedback heat the gas in and around the galaxies, limiting star formation and preventing further gas accretion to the systems \citep[e.g.,][]{Efstathiou1992, Bullock2000, Benson2002, Bovill2009, Wetzel2015, Sawala2016, Applebaum2021}. Galaxies with larger masses are expected to still form a significant fraction of their stars at early times, but are not expected to be quenched by reionization \citep[e.g.,][]{Christensen2024}. Instead, these systems should retain their gas, accrete additional material, and continue forming stars until the present-day unless stripped by external means (i.e., ram pressure or tidal stripping). 

When comparing to observations, the halo mass where quenching by reionization is thought to occur is often mapped to the more directly measurable quantity of stellar mass, with a typically quoted value of M$_*$ $\sim10^5$ \msun\ \citep[e.g.,][]{Bullock2017, Simon2019, WimberlyRodriguez2019}. As we are referring to the mass of galaxies that quench by reionization, the value of $10^5$ \msun\ is representative of the stellar mass both at the end of reionization {\em and} at the present-day. Thus, for convenience, we follow conventions in the literature and adopt a present-day stellar mass value of M$_*$ $\sim10^5$ \msun\ as the approximate upper limit where quenching by reionization is expected to be ubiquitous, which also provides a basis for comparison with empirical studies. Also note that, while there is considerable scatter in the {\em present-day} stellar mass-halo mass (SMHM) \citep[e.g.,][]{Nadler2020} and baryonic mass-halo mass relations \citep[e.g.,][]{McQuinn2022} at low galaxy masses (e.g., \mstar $\ltsimeq 10^7$ \msun), the SMHM relation is thought to have significantly less scatter at high-redshift (i.e., $z>4$) for galaxies with \mstar\ $\ltsimeq 10^{6.5}$ \msun\ \citep[e.g.,][]{Kim2024}, which makes the adoption of stellar mass as a tracer of the gravitational potential a reasonable approach.

The idea that very low-mass (\mstar\ $<10^5$ \msun) galaxies are quenched early is supported by observations of galaxies within the Local Group, where we have the largest sample of known faint systems and where we have been able to make the required detailed measurements. Hubble Space Telescope (HST) imaging of resolved stars to magnitudes below the oldest main sequence turn-off (oMSTO) has enabled the robust reconstruction of star formation histories (SFHs) across cosmic time. At photometric depths below the oMSTO, the age-metallicity degeneracy of the stars is broken and accurate ages can be assigned to the oldest stellar populations from a color-magnitude diagram \citep[CMD; e.g.,][]{Gallart2005}. The ensemble of the  stellar populations can then be used to infer a detailed history of the stellar mass assembly of the system. These SFHs reveal that nearly all of the galaxies below \mstar\ $\sim10^5$ \msun\ were quenched early, with the majority of their stellar mass formed in the first few Gyr after the Big Bang; this early quenching is often attributed to reionization \citep[e.g.,][]{Brown2014, Weisz2014, Skillman2017, Savino2023, McQuinn2023a}. 

However, this sample of galaxies is highly biased. Because of the sensitivity limitations of HST, most of these galaxies are currently satellites of the Milky Way (MW) or M31 and all of these systems are located within the Local Group. Their SFHs are intimately tied to their evolution in proximity of a massive host or in a group environment, making it difficult to separate the role reionization has played from local environmental processing. Note that simulations suggest the satellites of the MW were not necessarily (and unlikely) satellites of the MW at the time of reionization \citep{WimberlyRodriguez2019}. However, while the present-day location of a low-mass galaxy  does not necessarily represent the galaxy's nascent  environment, systems currently in the Local Group (particularly those at closer distances to the MW and M31) were likely formed in higher-density environments than present-day field galaxies. The idea that environment may be a dominant factor influencing low-mass galaxies is further reinforced in cosmological simulations. For example, low-mass galaxies as massive as \mstar\ $\sim10^8$ \msun\ and as distant as $\sim2 \times$ the virial radius of a massive galaxy are predicted to have their star formation quenched by environmental processing \citep[e.g.,][]{Fillingham2018}. Low-mass galaxies may be even more impacted by ram pressure stripping when near a pair of massive galaxies (like in the Local Group) due to increased gas density \citep{Samuel2022}. 

Thus, we assert that for the very low-mass galaxies near the MW or M31 in the Local Group it is not possible to unambiguously separate differences in the early mass assembly, unique patterns in SFHs, or any cessation or delay of star formation at ancient epochs between the intertwined effects of reionization, feedback, and ram pressure and tidal stripping in a group environment. In addition, the galaxies' orbital paths and infall times to the MW or M31, which are critical parameters for separating reionization from environmental effects, are highly uncertain, especially at early times \citep[e.g.,][]{Armstrong2021, Battaglia2022}. Existing studies indicate that quenching and differences in the SFHs in Local Group galaxies may be caused by many factors \citep[e.g.,][]{Weisz2015}.

For low-mass galaxies with \mstar\ $>10^5$ \msun, our empirical constraints are far coarser and contextualized by the dichotomy that low-mass galaxies in dense environments are predominantly quenched whereas field galaxies are predominantly star-forming (i.e., the morphology-density relation). However, there is evidence that when we are able to perform detailed studies of very low-mass, isolated galaxies, their histories deviate from expectations based on conclusions drawn from galaxies in closer proximity to the MW or M31, and, moreover, the SFHs disagree with most of the predictions. Note, in this regard, the case of the ultra-faint dwarf Pegasus~W. Pegasus~W has \mstar\ $= 6.5 \times10^4$ \msun\ but has indications of an extended SFH and thus would not have been quenched by reionization \citep{McQuinn2023b}. 

Furthermore, the few galaxies above the \mstar\ $\sim10^5$ \msun\ mass regime that lie in the outskirts of the Local Group (i.e., are considered isolated) and have the requisite data also have SFHs that deviate from general expectations. The Aquarius dwarf galaxy (\mstar\ $=1.6 \times10^6$ \msun) and Leo~A (\mstar\ $= 6.0 \times10^6$ \msun) both formed only a modest amount ($\ltsimeq$10\%) of their stellar mass at early times (i.e., within the first Gyr or so), followed by an extended period of quiescence post-reionization of several Gyr, with later re-ignition of star formation \citep{Cole2007, Cole2014}. Interestingly, the present-day satellite of M31 AndXVI (\mstar\ $\sim10^6$ \msun) shows a similar pattern in early star formation at a period when the galaxy was likely at a farther distance from M31. However, the star formation of AndXVI was quenched $\sim6$ Gyr ago, which has been speculatively attributed to environmental forces while the galaxy integrated into the M31 satellite system \citep{Monelli2016}. The SFH of the slightly more massive (\mstar\ $=4.3\times10^7$ \msun) isolated dwarf WLM derived from James Webb Space Telescope (JWST) NIRCam imaging shows a similar pattern \citep{McQuinn2024}, confirming earlier results from HST \citep{Albers2019}. Finally, there are hints that the lower mass galaxy Leo~T \citep[\mstar\ $= 1.4 \times10^5$ \msun;][]{McConnachie2012} also experienced extended periods of quiescence post-reionization and an overall slower build-up in stellar mass \citep{Clementini2012, Weisz2012}, although the larger uncertainties on the Leo~T SFH prohibit strong conclusions. At slightly higher masses, the effects of reionization are diminished -- for example, IC~1613 (\mstar\ $\sim$ 10$^8$ \msun) has a nearly constant SF \citep{Skillman2014}, as does the LMC \citep[\mstar\ $=3 \times 10^9$ \msun;][Cohen et al.\ in press]{Weisz2013, Ruiz-Lara2020, Mazzi2021}.

It has also been proposed that the extended or `slow' SFHs of the more isolated dwarfs in the Local Group reflect a deeper connection to their original environment \citep{Gallart2015}. In this scenario, due to the lower density of material at early times, the dwarfs assemble both their dark matter halo and their baryonic components on a slightly later timescale than low-mass systems located in more dense environments, which exhibit `fast' stellar mass assembly.

Regardless, this emerging SFH pattern for isolated, low-mass galaxies suggests that there may be a mass transition range where reionization can still impact the SFH without fully quenching activity and that, when we have the requisite quality data, the majority of stellar mass in low-mass galaxies is revealed to have formed relatively recently (i.e., $z<1$). While this SFH behavior has been noted to occur in simulations \citep{Wright2019}, it is not typically reproduced \citep{Fitts2017, Applebaum2021}. Instead, high-resolution cosmological simulations reveal a more rapid growth process. For example, a detailed comparison of over 100 low-mass simulated galaxies formed in a variety of environments, ranging from field galaxies in a Local Volume-like environment to systems near MW-mass halos, reveal clear but more subtle environmental trends in the SFHs \citep{Christensen2024}. The low-mass galaxies within 1 Mpc of a massive galaxy form their stellar mass over a shorter time period than those in the field (i.e., at greater distances from a massive galaxy). On the other hand, in these simulations the majority of even the isolated systems accumulate more than half of their stellar mass within the first few Gyr.

The paucity of the required observations of low-mass galaxies over a range in mass and in different environments limits our empirical understanding and prevents us from discerning between models. Clearly what is needed are SFHs of a sample of galaxies spanning the mass range of interest (i.e., $10^4$ \msun $<$ \mstar\ $<10^8$ \msun) that are {\em isolated} so we can separate the evolution of galaxies as a function of mass independently from their environment. 

The advent of the JWST \citep{Gardner2023, Rigby+2023}, with its greater sensitivity and resolution compared with HST, makes it now possible to obtain the deep imaging required for precisely measuring SFHs out to greater distances, allowing us to study galaxies in different environments. We have obtained such data on the galaxy \lp. Located just outside the Local Group (D $= 1.62\pm0.15$ Mpc), Leo~P is isolated, has a low stellar mass (\mstar\ $= 2.9\times10^5$ \msun; see Section~\ref{sec:stellarmass}), a comparable gas mass \citep[$M_{HI} = 8.1\times10^5$ \msun;][]{Giovanelli2013}, and is forming stars \citep{Rhode2013}. \lp\ is the prototypical, isolated field galaxy that lies in the parameter space where the predictions diverge in cosmological models, dark matter physics, and reionization models, thereby providing a unique laboratory to explore the early evolution of a low-mass galaxy in detail.

\subsection{A Summary of Leo~P's Properties: A Quintessential, Isolated, Low-Mass, Metal-Poor Galaxy}\label{sec:leop_props}
Leo~P was discovered via its neutral gas content in the Arecibo Legacy Fast ALFA (ALFALFA) \hi\ survey \citep{Giovanelli2013}. Follow-up Very Large Array \hi\ observations show an ordered gas rotation in \lp\ with an extremely low circular velocity \citep[$v_{circ} = 15\pm5$ km s$^{-1}$;][]{Bernstein-Cooper2014}. Ground-based BVRI and H$\alpha$ imaging confirmed the presence of young stars, a single bright H{\sc ii} region, and an extended older stellar population \citep{Rhode2013, McQuinn2013}. Large Binocular Telescope (LBT) optical spectroscopy including the [O III] $\lambda$4363 emission line yielded a ``direct'' oxygen abundance of 12 $+$ log(O/H) $= 7.17\pm0.04$; Leo~P is one of the most metal-poor star-forming galaxies in the nearby Universe \citep{Skillman2013} and one of the few known extremely metal-poor galaxies that has properties consistent with extrapolations of the mass-metallicity relation and the only one consistent with the luminosity-metallicity relation \citep{McQuinn2020}. The addition of LBT infrared (IR) spectroscopy enabled a high-precision measurement of the helium abundance, relevant to the determination of the primordial helium abundance \citep{Aver2021}. 

\lp\ hosts a single O star with an \hii\ region that approximates a Str{\"o}mgren sphere. VLT MUSE provided the first stellar spectroscopy of this extremely metal-poor (3\% \zsun), massive O star, and also provided spectacular mapping of the ionized hydrogen structures in the galaxy \citep{Evans2019}. Far UV spectra of the O star in \lp\ taken with HST Cosmic Origins Spectrograph (COS) were nearly devoid of any stellar wind signatures, but showed line broadening that suggests the star has a high projected rotational velocity \citep{Telford2021}. This high rotational velocity is consistent with the weak far UV wind features as winds are primarily responsible for the transfer of angular momentum away from a star and rotational spin-down \citep[e.g.,][]{Meynet2002, Groh2019}. Keck Cosmic Web Imager (KCWI) optical integral field unit spectroscopy of the \hii\ region enabled a measurement of the ionizing photon production rate of the star. These first empirical constraints on the ionizing properties of a massive star in the metallicity regime similar to stars thought to reionize the universe at early times compared favorably with expectations from widely used theoretical model stellar spectra \citep{Telford2023}. Similarly, candidate dust-producing AGB stars identified in coordinated Spitzer Infrared Array Camera (IRAC) and HST Wide Field Camera 3 (WFC3) infrared (IR) imaging provide an opportunity to constrain dust production in metal-poor high-redshift analog galaxies \citep{Goldman2019}. 

HST ACS imaging of the resolved stellar populations produced the deepest CMD of any galaxy outside the Local Group, reaching more than 2 mag below the red clump, which provided a secure distance, identified RR~Lyrae stars, and constrained the SFH of \lp\ \citep{McQuinn2015a}. One of the main results of this work was that, despite its low mass, \lp\ was not quenched by reionization or stellar feedback, and continues today as a star-forming galaxy. While the data that provided these constraints were deep given the distance to \lp\ and the sensitivity of HST, the imaging still did not reach the necessary depth to convincingly measure the ancient SFH of the galaxy, which is the focus of this work using deeper imaging enabled by the JWST. 

Finally, combined with the previously measured gas-phase oxygen abundance and gas content, the resolved stellar populations observations constrained the chemical evolution history of \lp\ and enabled a measurement of the production, distribution, and retention of oxygen in \lp. Based on this analysis, the galaxy has retained only 5$\pm2$\% of its oxygen produced by nucleosynthesis, with 4$\pm$2\% residing in the ISM and the remaining oxygen locked in stars and stellar remnants \citep{McQuinn2015b}. New HST COS FUV spectra, combined with existing ground-based optical spectra, will constrain the N/O and C/O relative abundances in \lp, furthering our understanding of the chemical evolution of galaxies at very low masses (Danielle A.\ Berg, in preparation). 

\begin{table}
\begin{center}
\caption{Properties of \lp\ and Observation Details}
\label{tab:properties}
\end{center}
\begin{center}
\vspace{-15pt}
\begin{tabular}{lr}
\hline 
\hline 
\multicolumn{2}{c}{\lp\ Properties} \\
\hline
\hline
RA (J2000) 				& 10:21:42.509 \\
Dec (J2000)				& $+$18:05:16.09 \\
$\mu$	(mag)				&  $26.05\pm0.20$ (1) \\
Distance (Mpc)			& 1.62$\pm0.15$ (1)\\
12$+$log(O/H) 	 		& 7.17$\pm0.04$ (2) \\
\mstar (\msun) 			& $2.7\pm0.4\times10^5$ \\
M$_{HI}$ (\msun)			& $8.1\times10^5$ (1; 3) \\
\hline
\multicolumn{2}{c}{Galactic Extinction} \\
\hline
$A_V$ (mag)				& 0.07 (4) \\
$A_{F090W}$ (mag)			& 0.03 (4) \\
$A_{F150W}$ (mag)			& 0.01 (4) \\
$A_{F277W}$ (mag)			& $<$0.01 (4) \\
\hline
\hline
\multicolumn{2}{c}{JWST NIRCam Observations} \\
\hline
\hline
PID					& JWST-GO-1617\\
F090W (s)				& 66,139 \\
F150W (s)				& 39,683 \\
F277W (s)				& 105,822 \\
\hline
\multicolumn{2}{c}{50\% Completeness Limits} \\
\hline
F090W (mag; \msun)	&  29.21; 0.76\\
F150W (mag; \msun)	&  28.85; 0.76 \\
F277W (mag; \msun)	&  28.55; 0.78 \\
\hline
\multicolumn{2}{c}{Timescales} \\
\hline
\tninety (Gyr)				&  1.90 $^{+0.15}_{-0.08}$ \\
\teighty (Gyr)				&  2.75 $^{+0.05}_{-0.05}$ \\
\tfifty (Gyr)				& 5.42 $^{+0.27}_{-0.16}$ \\
\ttwentyfive (Gyr)			& 9.52 $^{+0.13}_{-0.11}$ \\
\tten (Gyr)				&13.11 $^{+0.07}_{-0.24}$ \\
\hline           
\end{tabular}
\end{center}
\tablecomments{RA and Dec are the coordinates of the NIRCam pointing. References: (1) \citet{McQuinn2015a}; (2) \citet{Skillman2013}; (3) \citet{Giovanelli2013}; \citet{Schlafly2011}. 
The 50\% completeness limits are listed in magnitudes and the corresponding mass of a 14 Gyr, [M/H] $=-1.6$ star based on the PARSEC stellar library. Timescales are based on the SFH derived from the F090W-F150W CMD with the PARSEC library. }
\end{table}

\subsection{The Local Environment Around \lp: Assessing Isolation}\label{sec:environment}
\lp\ is at a distance of 1.62$\pm0.15$ Mpc, which places the galaxy outside the Local Group. Importantly, this means \lp\ is far enough from the MW and M31 that it does not experience ram pressure stripping from the hot halos of these galaxies, nor from any diffuse gas that may permeate the Local Group. It also means that \lp\ is not impacted by the strong tidal forces that low-mass galaxies in closer proximity to massive spirals experience. \lp's location outside the Local Group helps explain why, in contrast to nearly all galaxies of similar mass in the Local Group, \lp\ is gas-rich and star-forming. Indeed, the only galaxies within the Local Group in a similar mass regime (i.e., within $\sim1$ dex in \mstar\ of \lp) that are still gas-rich and star-forming are all found in the outskirts of the Local Group (i.e., Leo~A, Aquarius), or are on their first infall based on radial position and velocity \citep[i.e., Leo T;][]{Rocha2012, Bennet2023}. 

However, being outside a galaxy group environment does not guarantee that a galaxy is isolated. Low-mass galaxies are often found in loose associations of other low-mass galaxies, with varying inter-galaxy distances and unknown interaction histories. The association nomenclature dates back to \citet{Tully1988} and describes the spatial linkage between galaxies that have such insignificant luminosities that the luminosity density fails to reach the threshold to be called a galaxy group. 

\lp\ resides at the end of one such dwarf galaxy association, labeled 14$+12$ \citep{Tully2002, Tully2006}. 14$+12$ includes six known galaxies arranged with a fairly linear alignment stretching $\sim$1 Mpc with a width of only $\sim$100 kpc. The association includes, in order of distance from the MW, NGC3109, its two known satellites Antlia~B and Antlia, Sextans~A, Sextans~B, and \lp\ \citep{McQuinn2013}. 14$+$12 is sometimes referred to as the NGC3109 association as NGC3109 is the most massive of the galaxies. Given its present-day location at the end of this string of dwarfs, \lp\ appears isolated \citep[the distance to its nearest neighbor, Sextans~B, is $\sim$0.4 Mpc;][]{McQuinn2013} and is unlikely to have been significantly influenced by the other low-mass galaxies in the 14$+$12 association in the recent past. Furthermore, results from hydrodynamical simulations suggest that if any interaction occurred, it most likely was a merger at early times before $z \sim 2$ \citep[e.g.,][]{Fitts2018, Gandhi2023}. Given these circumstances, \lp's evolution seems likely to have been driven by secular processes, rather than by its environment. 

Even so, the origin and dynamical history of the 14$+$12 galaxy association may be relevant to understanding Leo~P's evolution. Based on a typical galaxy formation process in the $\Lambda$CDM paradigm, the association may have formed via collapse along a cosmic filament without any previous association with the Local Group. On the one hand, this would suggest that \lp\ has been generally isolated throughout cosmic time. On the other hand, the large recessional velocities measured for members of the 14$+$12 association (120$-$160 km s$^{-1}$) suggest that the galaxies may have experienced a previous fly-by interaction with the Local Group $\sim$7 Gyr ago ($z < 1$) which boosted their velocities \citep{Bellazzini2013, Shaya2013}. However, there are still significant uncertainties in the orbital modelling due to, e.g., uncertain masses of the MW, M31, and the Local Group as a unit \citep[e.g.,][]{Chamberlain2023}.  Alternatively, the galaxies in the association may even have once resided within the Local Group and could be backsplash galaxies at the present day \citep{Banik2021}. While the 14$+$12 association is generally considered unbound \citep{Kourkchi2017}, these latter two scenarios would imply that the galaxies were a bound group in the past \citep{Micic2022}, but have become unbound based on the interaction with the Local Group.

The present work on the ancient SFH of \lp\ can provide additional constraints on the above scenarios: since mergers, fly-bys, and previous interactions can trigger star formation in gas-rich galaxies via compression, they can leave a discernible imprint on the SFH if the data are of sufficient quality. In a future work we will combine the JWST observations presented here with HST ACS data obtained in 2013 \citep[HST-GO-13376;][]{McQuinn2015a} to measure the proper motion and 3-D velocity of \lp\ and reconstruct its orbital history. 

\vspace{10pt}
The paper is organized as follows. In Section~\ref{sec:data}, we describe the observations and data reduction and present the CMDs. In Section~\ref{sec:sfh}, we derive the SFH. In Section~\ref{sec:discuss}, we compare the SFH of \lp\ to those of three other isolated, low-mass galaxies and discuss the implications of our results. In Section~\ref{sec:conclusions}, we summarize our findings. For redshift calculations, we adopt a Planck cosmology \citep{Planck2018}. 

\begin{figure*}
\begin{center}
\includegraphics[width=0.46\textwidth]{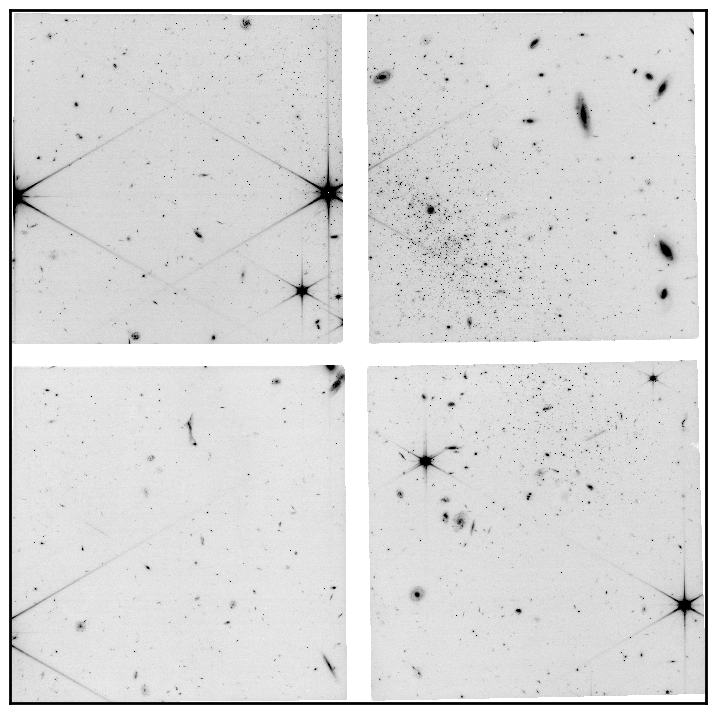}
\includegraphics[width=0.49\textwidth]{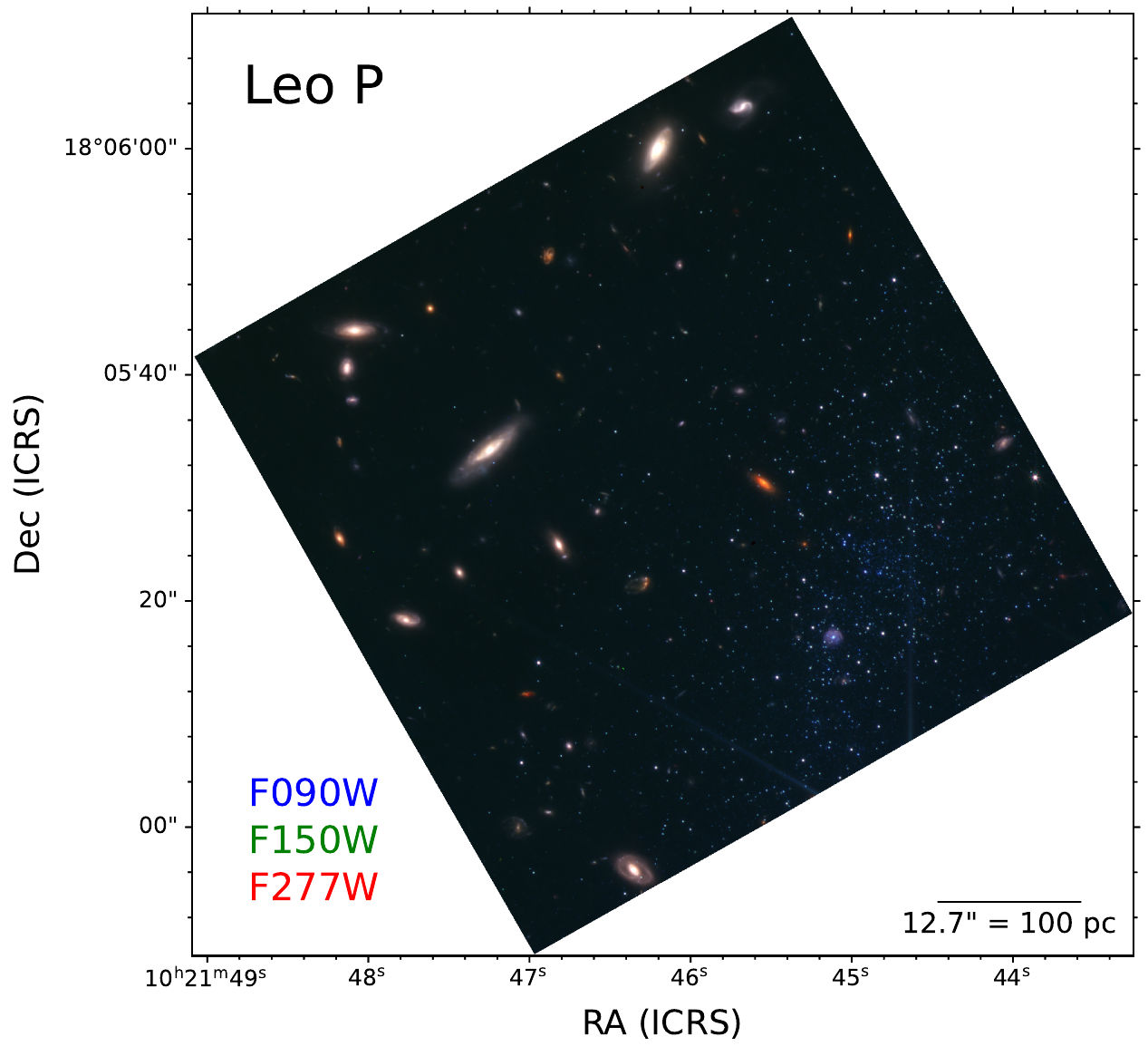}
\end{center}
\caption{Left: F090W image of the full field of view in the native orientation of the observations. Clockwise from the top left, the NIRCam detectors are B3, B1, B2, and B4. The center of the stellar disk falls on the B1 detector in the upper right of the figure. Right: Three color image zoom-in on B1 with the main stellar body of \lp. The image is created using the F090W (blue), F150W (green), and F277W (red) filters oriented N up and E left. Note the rich range in object colors including the redder background galaxies and the bluer colors of the stars in \lp.}
\label{fig:image}
\end{figure*}

\section{Observations and Data Processing}\label{sec:data}
\subsection{Observing Strategy}
\lp\ was imaged with the NIRCam instrument \citep{Rieke2023} for 35.8 total hours between April  8-12, 2023. The imaging was obtained in two visits, but with the same {\sc orient} ensuring accurate alignment of all images for precise, simultaneous PSF photometry, and improved sampling of short period variable stars. The observations were taken with the standard sub-pixel dither pattern with seven dithers. The dithers help to mitigate against hot pixels and improve flat fielding thereby increasing the signal-to-noise ratio (S/N) reached in the total science times and improving sampling of the PSF. We opted not to dither to cover the detector gaps as uniform depth on the imaged stellar populations is valued over complete areal coverage of the stars and requires less overhead time. The stellar disk of \lp\ is covered by the four detectors in one of the NIRCam modules. Thus, we elected to use just module B for the observations, enabling a large number of exposures while keeping the observations under the data rate and data volume limits. 

The observations include imaging with the F090W and F150W filters in the short wavelength (SW) channel and the F277W filter in the long wavelength (LW) channel. We chose the F090W and F150W filters for our primary science objective of reconstructing the SFH of \lp\ as the F090W - F150W combination has a wide enough color baseline to accurately separate stars in different phases of stellar evolution and the SW filters have a smaller PSF compared to the LW channel. We opted for simultaneous imaging with the F277W to explore the efficacy of recovering a SFH using the F277W filter paired with one of the SW filters. We found the star recovery in the F277W filter sub-optimal for SFH recovery work, which we discuss further below.

The observing program was designed to reach a photometric depth below the oMSTO in the F090W $-$ F150W filters ($M_{F090W} = +3.5$; $M_{F150W} = +2.7$) with a S/N of 10 to enable the robust reconstruction of the ancient SFH. To reach our desired S/N at these depths, we used the {\tt DEEP8} readout pattern to keep the data rate below the limits and chose five groups per integration. We then optimized the number of integrations and dithers to reach our required S/N and depth while also minimizing overhead and maintaining integration times per exposure below the 10000~s limit. This resulted in 10 integrations per exposure for F090W and six integrations per exposure for F150W with seven dithers. The observing strategy resulting in a science time of 18.37 hours in F090W and 11.02 hours in F150W. 
As the F277W filter was used as the LW complement to both the F090W and F150W exposures, the total science time in F277W was 29.39 hours. The breakdown of time per filter is also listed in Table~\ref{tab:properties}. 

\subsection{Images}
The left panel in Figure~\ref{fig:image} presents the full F090W image in an inverted grayscale and in the native orientation of the NIRCam observations. While difficult to discern in the grayscale rendering, we note the presence of diffuse stray light features (i.e., `wisp' artifacts) that are present in all NIRCam exposures in the same B3 and B4 detector locations,\footnote{For details, see JDox \citep{2016jdox}; \url{https://jwst-docs.stsci.edu}.} 
which are brightest in the F150W filter (not shown). These do not impact our photometry given the distance from the main stellar body of \lp. Furthermore, our photometry approach subtracts a local sky background for each point source and would account for the presence of this artifact. The right panel shows a three color image of the B1 detector which contains the main stellar body of \lp. The galaxy is faint but visible and the stars have notably blue colors compared to the redder background galaxies in the image given our adopted color palette. 

\begin{figure}
\begin{center}
\includegraphics[width=0.37\textwidth]{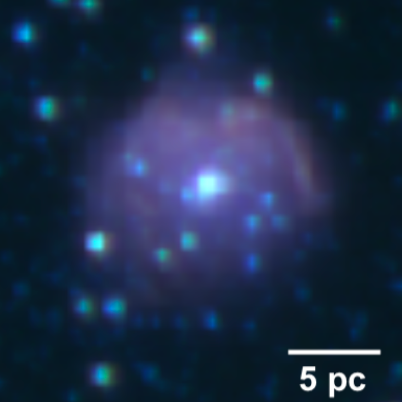}
\end{center}
\caption{Zoom-in on the single \hii\ region in \lp. Individual well-recovered sources that are co-spatial within the \hii\ region are marked in the CMDs in Figure~\ref{fig:cmds}. The image is oriented N up and E left with a vertical extent of 2.8\arcsec\ corresponding to $\sim22$ pc at the distance of \lp.}
\label{fig:hii_image}
\end{figure}

\begin{figure*}
\begin{center}
\includegraphics[width=0.48\textwidth]{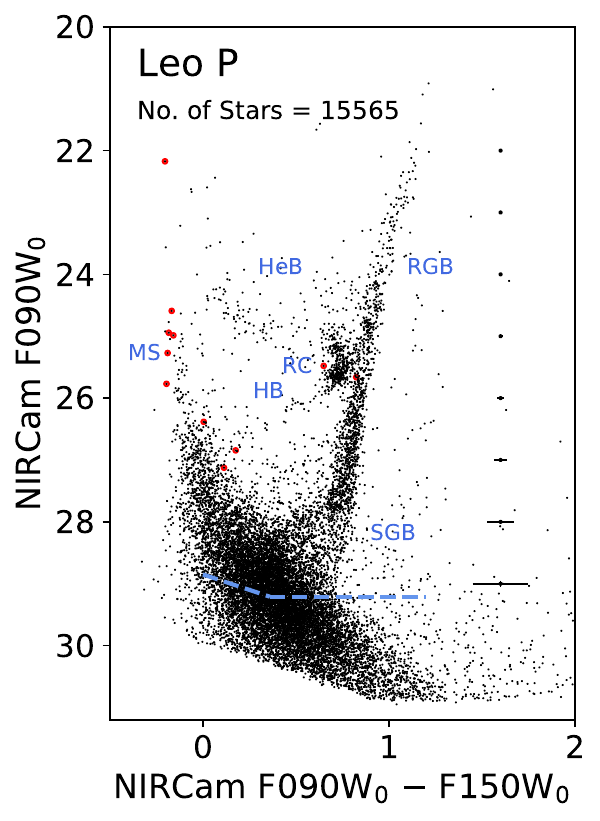}
\includegraphics[width=0.48\textwidth]{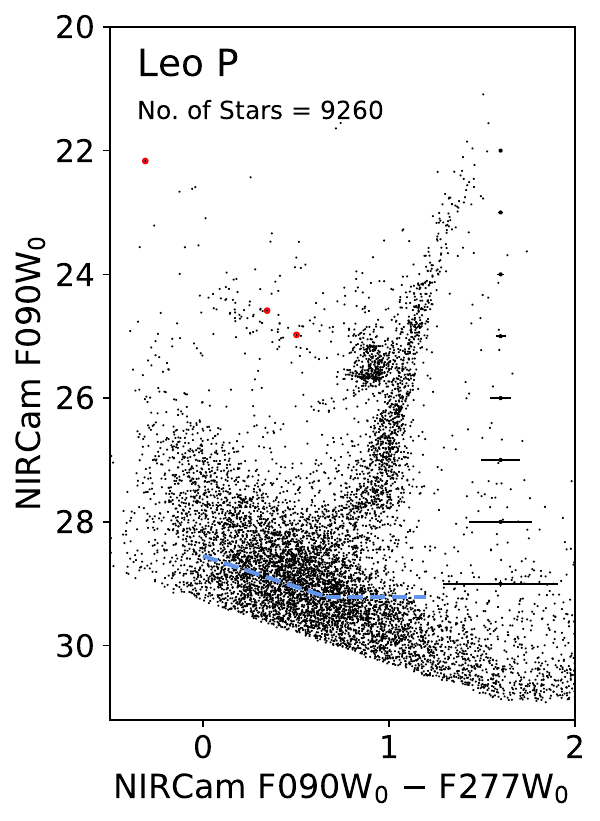}
\end{center}
\caption{CMDs created from our final stellar catalogs for \lp\ with the F090W, F150W filter pair (left) and F090W, F277W filter pair (right). In the left panel we mark the regions for the main stages of stellar evolution including the upper main sequence (MS), young blue and red helium burning (HeB) stars, the red giant branch (RGB), the red clump (RC), the horizontal branch (HB) and the sub-giant branch (SGB). The eleven sources that are co-spatial with the \hii\ region within a radius of 0.72\arcsec are marked in red. The brightest of these sources located at the top of the main sequence powers the \hii\ region. Only three sources in the same region are recovered in the F277W photometry due to the increased impact of crowding in the LW data. Representative uncertainties are shown on the right in each CMD and the 50\% completeness limits determined from artificial star tests are drawn as dashed blue lines. The F150W data are deeper than the F277W data, and the number of point sources recovered in the F090W-F150W catalog is more than 50\% greater than the number in the F090W-F277W data. See Section~\ref{sec:cmd} for details.}
\label{fig:cmds}
\end{figure*}

Figure~\ref{fig:hii_image} presents a zoom-in on the single \hii\ region in \lp\ using the same color scheme as in Figure~\ref{fig:image}. The image spans $\sim22$ pc on a side and clearly conveys the detailed structure of the \hii\ region. As noted previously based on HST data, the \hii\ region is an excellent example of a Str{\"o}mgren sphere \citep{Telford2023}, while the higher resolution NIRCam imaging captures more detailed structure including a region of higher surface brightness nebular emission to the northwest.

\subsection{Photometry}
We performed point-spread function (PSF) photometric reductions of our observations using the software {\tt DOLPHOT} with the new NIRCam module \citep{Dolphin2000, Weisz2024}. These data were obtained from the Mikulski Archive for Space Telescopes (MAST) at the Space Telescope Science Institute. The specific observations analyzed can be accessed via \dataset[DOI: 10.17909/faz3-2616]{https://doi.org/10.17909/faz3-2616}. All images were processed by the JWST pipeline {\sc cal\_ver}=1.12.5, {\sc crds\_ver} = 11.16.20, and {\sc crds\_ctx} = jwst\_1174.pmap. We used the F090W level 3 {\tt i2d.fits} drizzled image as the astrometric reference frame for {\tt DOLPHOT} and performed photometry on the individual level 2 {\sc cal.fits} frames. The NIRCam zeropoints in {\tt DOLPHOT} are based on the Sirius spectrum.

The images were pre-processed before running photometry. We converted units from MJy/sr to data numbers (DN) using the appropriate JWST exposure time FITS keyword ``TMEASURE"\footnote{For details on the measurement time definition, see JDox;\\\url{https://jwst-docs.stsci.edu/accessing-jwst-data/jwst-science-data-overview/jwst-time-definitions\#JWSTTimeDefinitions-meastimeMeasurementtime}.} and masked out bad-pixels, identified saturated pixels, and applied pixel area maps using the {\tt DOLPHOT} routine {\tt nircammask}. For the photometry, we used model PSFs generated with the simulation tool WebbPSF Version 1.2.0, which accounts for charge diffusion effects, interpixel capacitance, and post-pixel coupling effects. Whereas the previous WebbPSF Version 1.1.0 produced PSFs that were too sharp relative to observations (i.e., the modelled PSF shape was too centrally concentrated), the updated models more accurately reproduce the observed PSF, thereby improving the precision of our photometry \citep{Weisz2024}.

\begin{figure*}
\begin{center}
\includegraphics[width=0.95\textwidth]{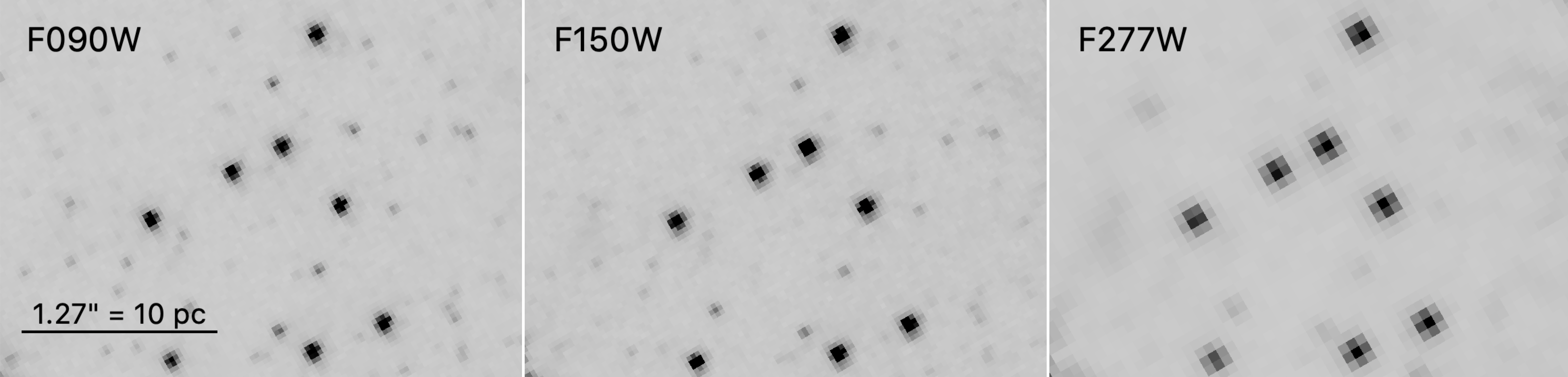}
\end{center}
\caption{Comparison of the image quality and sensitivity in the three filters using a matched field of view. Immediately apparent is the higher resolution at bluer wavelengths, with some sources that are clearly resolved in F090W becoming blended in the F277W filter. Closer inspection also reveals greater sensitivity as one moves from red to blue filters, which we quantify using artificial star tests (see Table~\ref{tab:properties}).}
\label{fig:zoom_filters}
\end{figure*}

The {\tt DOLPHOT} software includes an extensive list of parameters that the user can customize depending on the data and science objectives. We adopted the values from the JWST ERS program determined from comprehensive tests of the parameters using data of similar quality and for similar purposes \citep{Weisz2023, Weisz2024}, which are the same parameters recommended for ACS images for the SW filters and for WFC3/IR for the LW filters from the PHAT program \citep{Williams2014}. The parameters include {\tt FitSky} $=2$ for all images and {\tt RAper} $=2, 3$; {\tt Rchi} $=1.5, 2.0$; {\tt Rsky} $=(3,10), (4,10)$ for the SW and LW filters respectively. 

We experimented with running photometry just on the SW filters (F090W, F150W) versus including all three filters (F090W, F150W, F277W) simultaneously. We found that the SW+LW combined imagery resulted in fewer detections than SW only, which has previously been noted when running {\tt DOLPHOT} on ACS+WFC3/IR images. Thus, in order to maximize the source detection across the SW+LW data, we implemented a multi-step photometry process. First, we performed photometry on just the SW data using the F090W {\sc i2d.fits} file as the reference image. Second, we used the detection list resulting from this SW only run and the {\tt warmstart} option in {\tt DOLPHOT} to force photometry to be performed at all coordinates in the detection list. As a consistency check on the photometry runs, we compared the F090W, F150W output from the SW-only run with that from the SW+LW run and found they were in good agreement in terms of source distribution in the CMD.

The final SW+LW photometry output was filtered for well-recovered point sources based on a number of quality metrics. Specifically, we selected sources with a S/N $\geq$ 4 in each filter, {\sc error flag} $\leq$ 2, {\sc object type} $\leq$ 1,  {\sc sharp}$^2 \leq$ 0.0225 per filter, {\sc crowd} $\leq$ 0.5 for the F090W and F150W filters, and {\sc crowd} $\leq$ 1.5 for the F277W filter. We opted for a higher crowding value for the F277W output as this filter has a larger PSF than the SW filters \citep[the F277W PSF Full-Width Half-Maximum (FWHM) is $\sim1.8\times$ greater than that of F150W and $\sim2.9\times$ that of F090W; see JDOX ][]{2016jdox}; the final value of 1.5 mag was guided by the distribution of crowding values as a function of magnitude. Our final choice of quality cuts balances the purity of the stellar catalogs with completeness, erring on the side of lower completeness to reach slightly greater photometric depths. 

We ran $\sim500$k artificial star tests (ASTs) on the images to measure the photometric bias and completeness of the images. Artificial stars were injected into the images following the spatial distribution of the full photometric output and then recovered using {\tt DOLPHOT} with the same set-up used for the photometry. We applied the same quality cuts we used for the photometry to the AST outputs, marking any source that did not meet our requirements as `unrecovered'. From the ASTs, we measure the 50\% completeness limits in each filter and provide those values in Table~\ref{tab:properties} in apparent magnitudes and the mass of an old, metal-poor star based on the PARSEC stellar isochrones at those magnitudes \citep{Bressan2012} after adopting the distance to \lp\ from \citet{McQuinn2015a}.

\subsection{Color-Magnitude Diagrams}\label{sec:cmd}
Figure~\ref{fig:cmds} presents the F090W-F150W CMD (left) and the F090W-F277W CMD (right) for the sources that meet our photometric quality requirements for \lp. The 50\% completeness limits are plotted as dashed lines. Representative uncertainties per magnitude are shown to the right of the CMD and include uncertainties from the photometry and uncertainties determined from the artificial star tests. The photometry was corrected for the small amount of foreground Galactic extinction along the line of sight; values are listed in Table~\ref{tab:properties}. 

\begin{figure*}
\begin{center}
\includegraphics[width=0.48\textwidth]{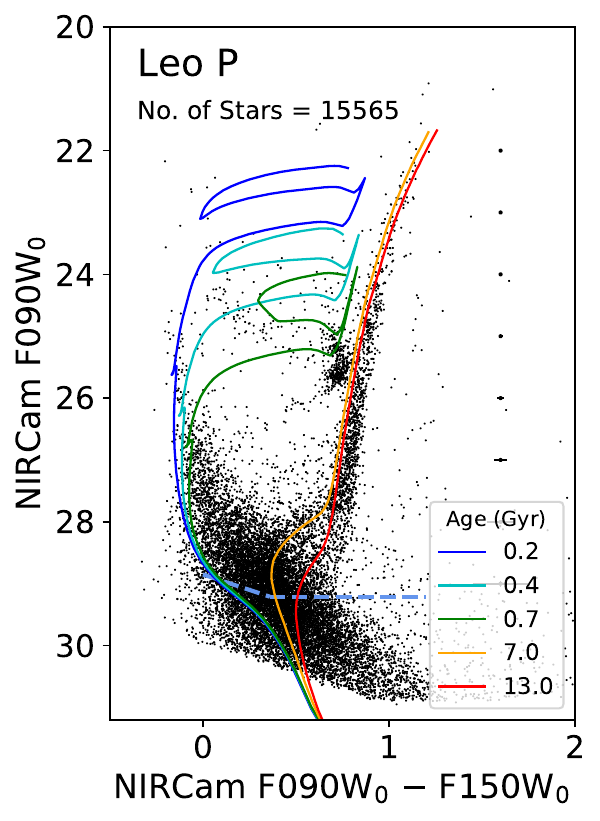}
\includegraphics[width=0.48\textwidth]{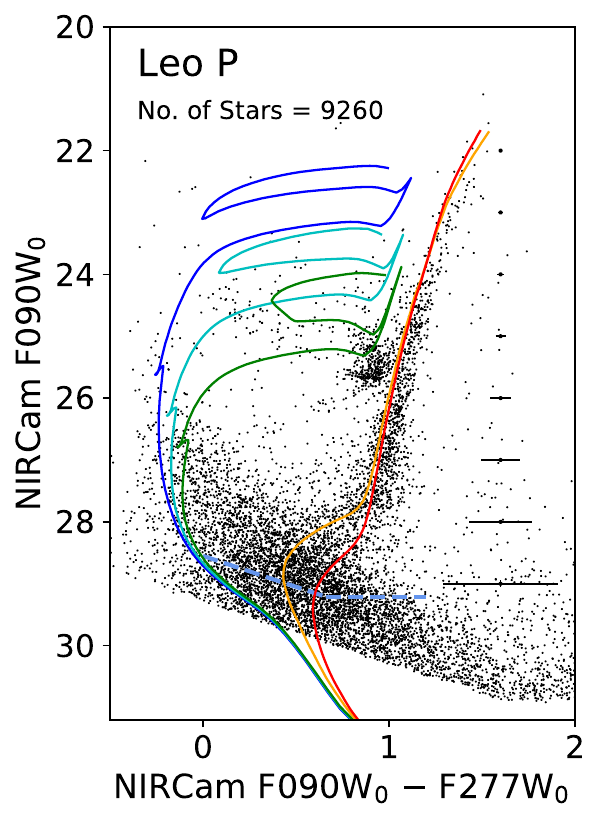}
\end{center}
\caption{The same CMDs as shown in Figure~\ref{fig:cmds}, now with PARSEC isochrones overlaid. The younger isochrones (0.2, 0.4, 0.7 Gyr) have [M/H]$= -1.4$ while the older isochrones (7, 13 Gyr) are slightly more metal poor ([M/H]$= -1.6$). }
\label{fig:cmds_isochrones}
\end{figure*}

Focusing on the F090W-F150W CMD, the photometry is exquisite and exceptionally deep, reaching approximately a magnitude below the oMSTO. This is the first time this photometric depth for resolved stars has been achieved for any galaxy outside the Local Group. The F090W-F277W CMD reaches below the oMSTO, but not to the depth of the F090W-F150W data. The difference in the F277W and F150W filters results in fewer total sources recovered in the F090W-F277W CMD ($\sim9.3$k) compared with the F090W-F150W CMD ($\sim15.6$k). The lower star recovery reflects a number of differences between the F277W and F150W filters including: (i) the larger PSF at the longer wavelength, which reduces the ability to recover stars in crowded regions, especially in the center of \lp\ and for stars at fainter magnitudes, (ii) the deeper data in the F150W data where additional stars are recovered, and (iii) the challenges in recovering bluer stars on the lower main sequence in F277W where the longer wavelength falls farther on the Rayleigh-Jeans tail of the spectral energy distributions. 

Figure~\ref{fig:zoom_filters} demonstrates these effects. From left to right, we show a small region of the data going from the shorter to longer wavelength filters in a matched field of view. Immediately apparent as one moves from the F090W image to the F277W image is the reduced ability to resolve individual sources and distinguish them from their neighbors and the lower S/N for faint sources. The difference in S/N moving to the F277W filter is especially striking given that the exposure time in this LW filter is  the sum of the exposure times in the two F090W and F150W SW filters from the simultaneous imaging (i.e., 106 ks in F277W vs.\ 66 ks in F090W and 40 ks in F150W). 

In Figure~\ref{fig:cmds}, stars in different stages of stellar evolution are readily identifiable. We mark a number of the corresponding features in the F090W-F150W CMD. Specifically, \lp\ hosts stars on the upper main sequence (MS) and a small number of blue and red helium burning (HeB) stars, which are all indicative of recent star formation. The red clump (RC) is well-defined and has a narrower width in the F090W-F150W colors compared with the F090W-F277W combination. The red giant branch (RGB) and horizontal branch (HB) are seen in both, but present as narrower sequences in the F090W-F150W CMD; this is likely due to increased photometric accuracy with the smaller PSF relative to the F277W filter. 

Finally, at faint magnitudes, the stars populating the sub-giant branch (SGB) have a non-uniform distribution in both CMDs. At brighter and bluer magnitudes, this feature is well-populated, whereas at slightly fainter and redder magnitudes the density of sources in the CMD is notably lower before increasingly slightly. This difference is difficult to discern in the F090W-F277W CMD as the S/N is lower in the F277W photometry at these faint magnitudes and the larger photometric uncertainties introduce considerable scatter in the CMD.

The different populations seen in the sub-giant branch correspond to stars of different ages. Figure~\ref{fig:cmds_isochrones} re-plots the CMDs, now with a series of isochrones from the PARSEC stellar library that have been updated with the in-flight JWST NIRCam filter transmission curves and the zero-points based on the spectrum of Sirius. The morphology of the higher density of sources at brighter magnitudes on the sub-giant branch is well-matched with a metal-poor ([M/H] $=-1.6$), 7 Gyr isochrone, whereas the higher density of sources at fainter magnitudes on the sub-giant branch is a better match to an older isochrone of 13 Gyr with the same metallicity. The two more prominent populations separated by a lower density of sources suggests there were two stronger star formation events $\sim13$ and $\sim7$ Gyr ago with an intervening period of lower star formation activity. Indeed, we recover this pattern in the SFH and will discuss this further in Section~\ref{sec:sfh_results}. 

We also overlay isochrones of younger ages in Figure~\ref{fig:cmds_isochrones}. Specifically, we plot isochrones for stars at a slightly higher fixed metallicity of [M/H] $=-1.4$ with ages of 200, 400, 700 Myr. These isochrones give an indication of the ages of the blue and red helium burning star sequences. Note, also, that the isochrones follow the red giant branch at brighter magnitudes, with some offset to bluer colors on the lower RGB. An offset to bluer colors is also seen on the lower main sequence for younger stars where the models are not a perfect match to the CMD.

\begin{figure*}
\begin{center}
\includegraphics[width=0.48\textwidth]{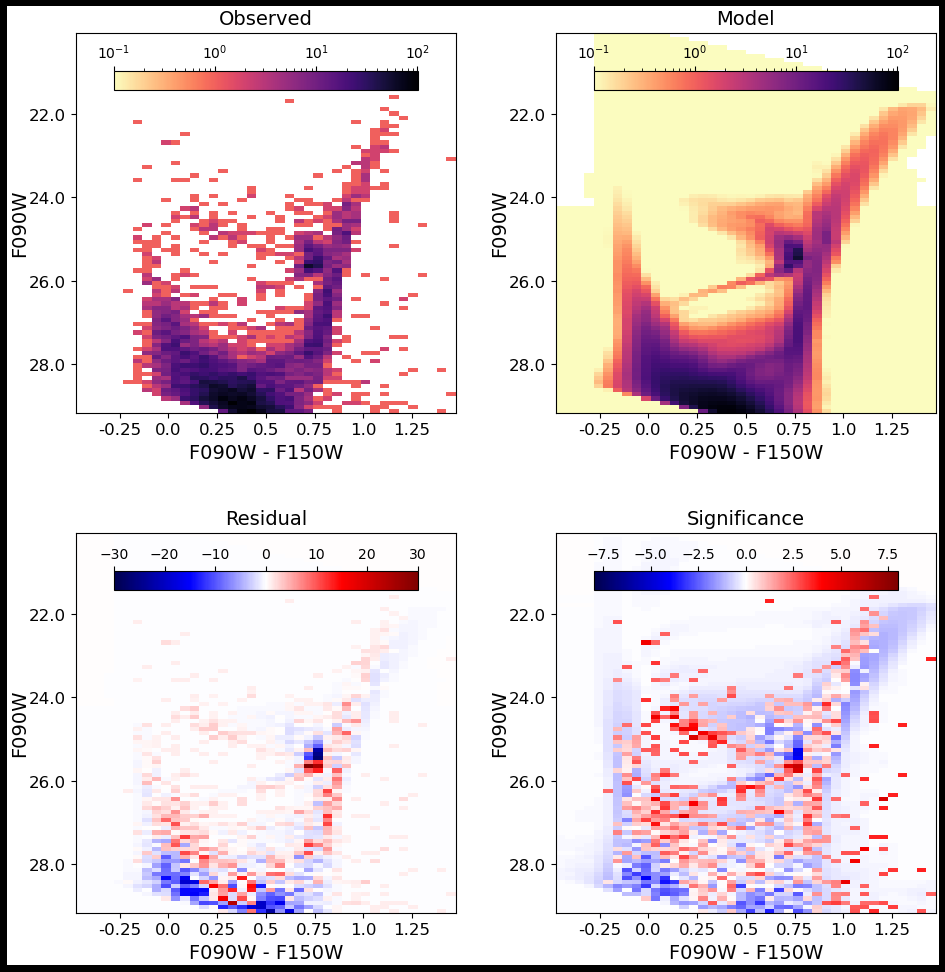}
\includegraphics[width=0.48\textwidth]{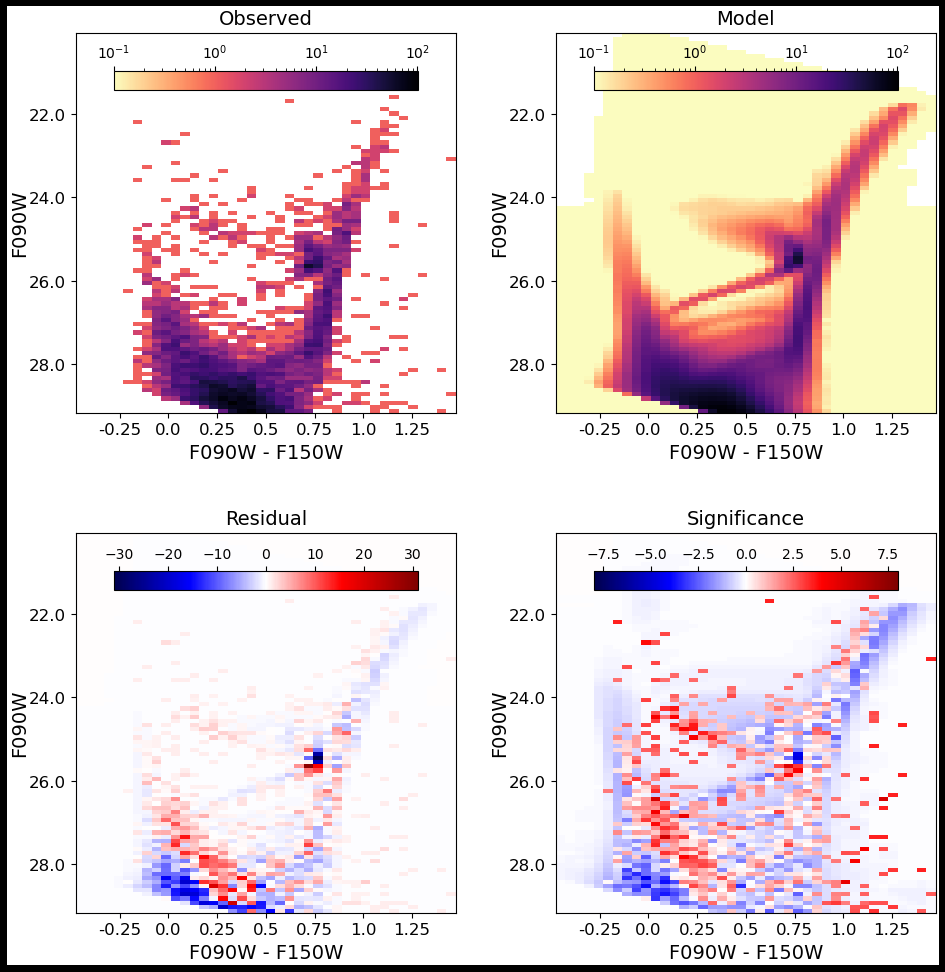}
\includegraphics[width=0.48\textwidth]{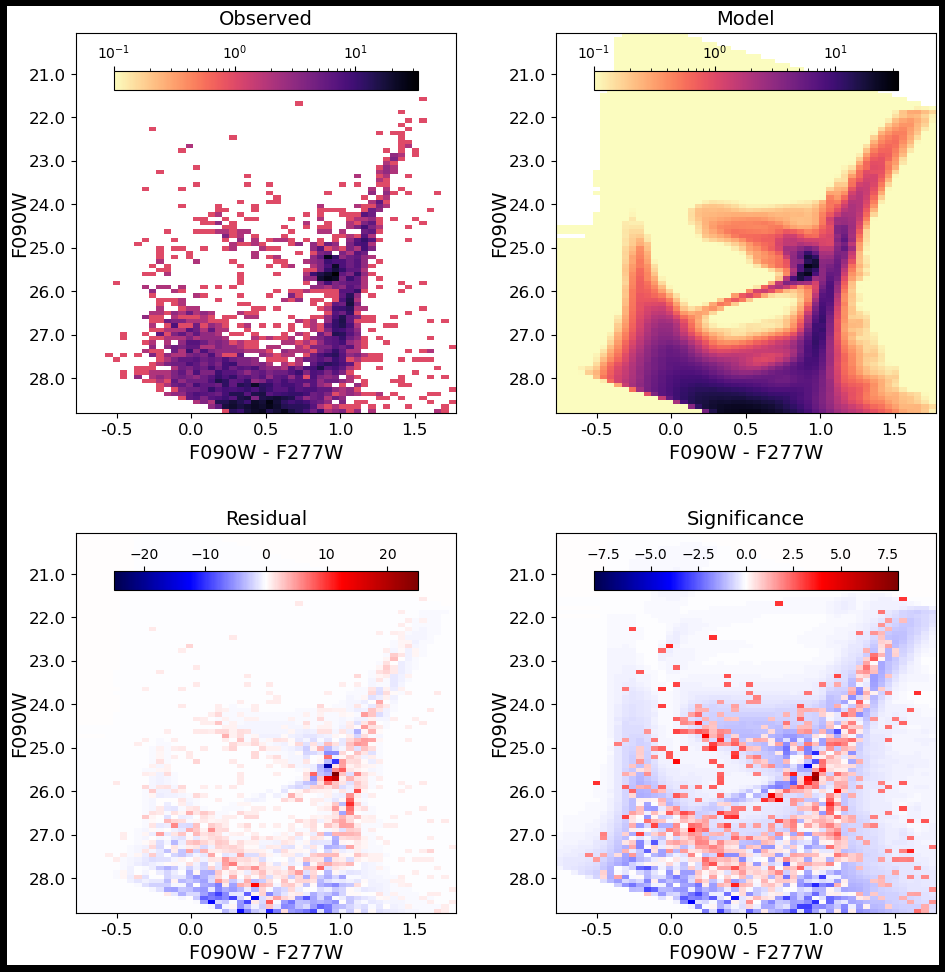}
\includegraphics[width=0.48\textwidth]{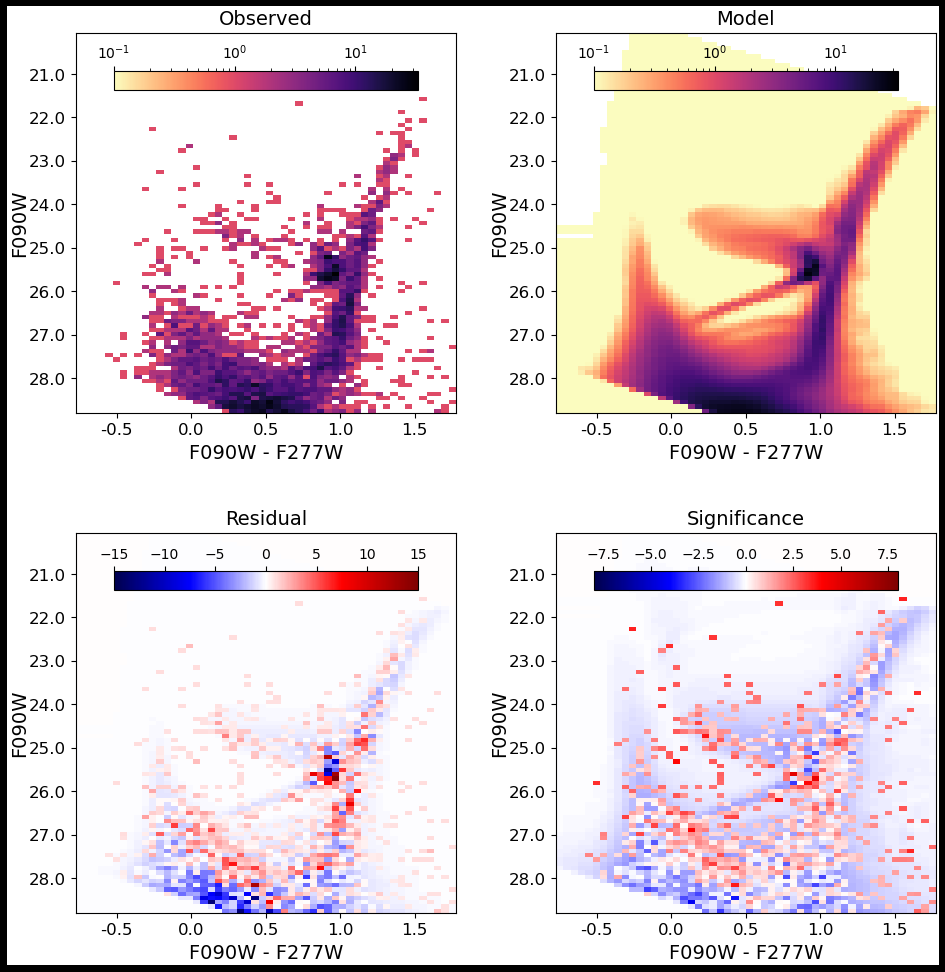}
\end{center}
\caption{A series of four-panel plots showing the CMD-fits for the different filter combinations with two different stellar libraries. The top two sets of plots are for F090W-F150W with PARSEC (left) and BaSTI (right). The bottom two sets of plots are for F090W-F277W with PARSEC (left) and BaSTI (right). Within each four panel plot we show: the observed CMD as a 2-D histogram or Hess diagram (top left), the modelled CMD as a Hess diagram (top right), the difference between the observed and modelled CMDs, where red (blue) indicates more (fewer) stars observed than in the model (bottom left), and the significance of the residuals or the observed $-$ model weighted by the variance in each Hess bin (bottom right). The scale of these residual significance plots are set uniformly from a minimum of $-$8 to a maximum value of 8. Overall, the modelled CMDs using the PARSEC and BaSTI libraries provide excellent fits to the observed NIRCam CMDs, with offsets seen most notably on the blue-side of the lower main sequence and the red-side of the upper RGB (see, also, Figure~\ref{fig:cmds_isochrones}).}
\label{fig:residuals}
\end{figure*}

\section{The Star Formation History of Leo~P}\label{sec:sfh}
\subsection{SFH Methodology}\label{sec:sfh_method}
The SFH was measured using the well-established technique of forward-modelling the CMD with stellar evolution libraries until the closest match to the observed CMD is found \citep[e.g.,][]{Tolstoy2009}. We use the CMD-fitting software {\sc match} \citep{Dolphin2002} which has been updated to fit JWST data. {\sc match} generates synthetic photometry of simple stellar populations with different ages and metallicities based on a user-specified stellar evolution library, a set of galaxy-specific parameters, and an assumed initial mass function (IMF). These synthetic CMDs are convolved with the photometric uncertainties and completeness function determined from the ASTs, combined with different weights, and iteratively compared to the observed CMD until the best fit is found using a Poisson likelihood function. The best-fitting modeled CMD encodes the most likely SFH of the galaxy. 

When fitting the data, we assumed a Kroupa IMF \citep{Kroupa2001} and a binary fraction of 0.35 with flat secondary mass ratio distribution. We adopted the distance to \lp\ of 1.62 Mpc from \citet{McQuinn2015a}. We experimented with re-fitting for the SFH using small perturbations on the distances and found the SFH solution was robust to small changes in the assumed distance. We also adopted a foreground extinction value of $A_V=0.073$ based on the dust maps of \citet{Schlegel} with a recalibration from \citet{Schlafly2011}. The extinction values corresponding to imaging in each NIRCam filter are listed in Table~\ref{tab:properties}. Given that \lp\ has a metallicity of only 3\% the solar value \citep{Skillman2013}, we expect little dust or extinction in the galaxy. Indeed, measurements derived from optical spectroscopy of the \hii\ region are consistent with an $A_V = 0.0$ \citep{Telford2023}. Thus, we assume no internal extinction is present.

The SFH solutions were derived using an age grid of log(t)= 6.6 $-$ 10.15 with time steps of log($\delta$t) = 0.1 dex for ages less than log(t)= 9 and log($\delta$t) = 0.05 dex for older ages,\footnote{Note that this results in finer {\em linear} time steps at more recent ages and coarser time steps at older lookback times.} and with a metallicity grid [M/H]= $-$2.0 to $-$0.9 and a resolution of 0.15 dex. Based on the reasonable assumption that galaxies become more chemically enriched with time, we required that the metallicity monotonically increase with time and placed a prior constraint on the present-day metallicity of [M/H] $= -0.9$. This upper limit was set to be higher than the spectroscopically measured present-day gas-phase oxygen abundance of 12+log(O/H) $= 7.17\pm0.04$ \citep{Skillman2013}, which is equivalent to 3\% of the solar value \citep{asplund2021} and the stellar metallicity of 2\% of Z$_\odot$ inferred for the massive O star (LP26) in \lp\ estimated by fitting its spectral energy distribution \citep{Telford2021}. The approach of requiring the age-metallicity relation (AMR) to increase in time has been shown to yield a metallicity distribution function (MDF) for stars that is consistent with a spectroscopically determined MDF \citep{McQuinn2024}. Nonetheless, we also tested the impact of allowing the AMR to be fit without constraints. The resulting AMRs have similar ranges to the results based on requiring an increasing chemical enrichment solution, but do show variations on the order of a few tenths of a dex in [M/H] in different time bins. These variations could be physical (i.e., the accretion of pristine gas at a specific epoch could lower the average metallicity of stars formed at that time), or they could simply be due to a slightly preferred fit to the models based on more degrees of freedom that are not physical. Importantly, the SFH solutions based on an unconstrained AMR are in agreement with results based on requiring the AMR to increase with time. Thus, our main results are not impacted by the choice in parameterizing of the AMR.

The SFHs presented here are based on two stellar evolution libraries, namely PARSEC \citep{Bressan2012} and BaSTI \citep{Hidalgo2018}. We chose these libraries because both have been updated with NIRCam in-flight filter transmission curves, Sirius zeropoints, and the absolute flux calibration from September 2023 (Martha L.\ Boyer et al., in preparation). The differences in the SFH solutions from the two stellar libraries give an indication of how the different assumptions used in the stellar models impact our results, which provides an initial estimate of the systematic uncertainties due to our imperfect understanding of stellar evolution. However, with only two libraries, these differences likely underestimate the true systematic uncertainties of the solutions to some degree. A full measure of the systematic uncertainties will be performed in future work once additional stellar libraries have been updated, which will also enable estimating uncertainties using Monte Carlo simulations that are tuned based on solutions from multiple stellar libraries \citep[e.g.,][]{Dolphin2012}. Statistical uncertainties due to the finite number of stars in the CMD were estimated using a hybrid Markov Chain Monte Carlo approach \citep{Dolphin2013}. 

Our main results focus on the SFH fits using the F090W and F150W data as the photometry is deeper and of higher quality compared to the photometry from the F277W imaging. However, as an exploratory exercise, we also independently fit the F090W-F277W CMD. We applied the same assumptions and methodology that were used for the F090W-F150W stellar catalog and present a comparison of results to the F090W-F150W fits below. 

\subsection{Best-Fitting SFH}\label{sec:sfh_results}
Figure~\ref{fig:residuals} shows the quality of the SFH fits in a 4-panel Hess diagram format (i.e., a 2-D histogram presentation of a CMD) based on the PARSEC (left columns) and the BaSTI (right columns) libraries using the F090W-F150W CMD (top panels) and F090W-F277W CMD (bottom panels). Within each 4 panel sub-plot, the top row shows the observed and the modeled CMD and the bottom rows shows the difference between the observed and modeled Hess diagrams and the residual significance (i.e., the observed $-$ model weighted by the variance in each Hess bin). The most informative panel to assess the overall fit is the residual significance plot in the lower right. A checkerboard pattern indicates that the data are well-fit by the model; red colors indicate regions where the model has too few stars relative to the observed CMD, whereas blue colors indicate an over-prediction of the number of stars. Overall, the modeled CMDs are well-matched to the data, with no clear trends seen in the weighted residuals. The main exception is the red clump, which is known to be problematic in both optical and near-infrared CMDs \citep[e.g.,][]{Gallart2005}. We also note a smaller mismatch in color in the RGB and the blue edge of the lower main sequence where the models have difficulty reproducing the full detailed structured of the data. The significance in the residuals in both these regions of the CMD varies between model and filter combination, and is most pronounced in both cases for the PARSEC fit to the F090W-F150W CMD. While it is difficult to directly assess the impact of these residuals on the SFHs recovered for \lp, the significance in all the fits is still quite small (mostly $<3$ standard deviations), and therefore unlikely to have a significant impact on the recovered SFH and stellar mass estimates. Note, also, that similar mismatches between model and data were previously reported in the fit to the NIRCam CMD of WLM \citep{McQuinn2024}. In this work, the SFH fits from the NIRCam F090W-F150W CMD were found to be in very good agreement with the SFH fits from ACS F475W-F814W CMD based on stars from the same region of the WLM, despite the fits showing differences in the best-fitting model NIRCam CMDs similar to our fits to \lp. These results supports our conclusion that the differences do not significantly impact the SFH results.

\begin{figure*}
\begin{center}
\includegraphics[width=0.98\textwidth]{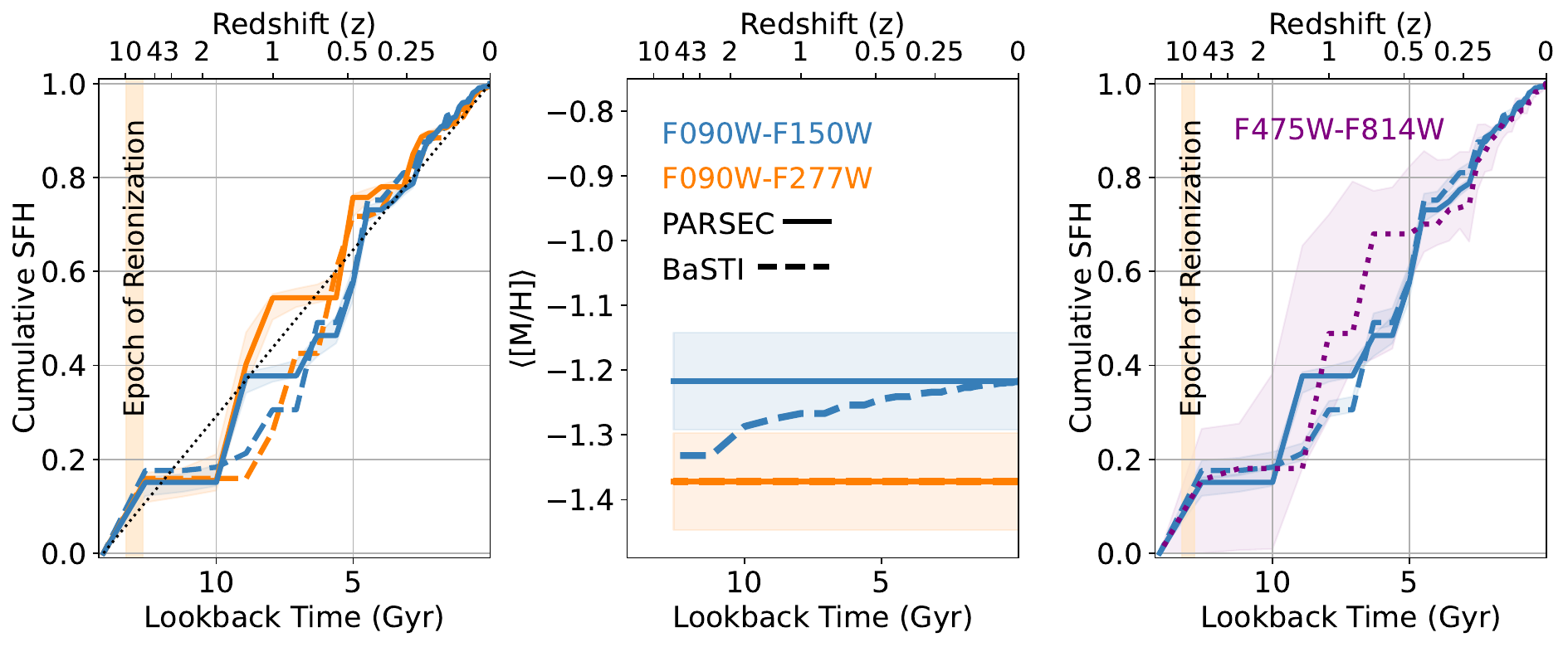}
\end{center}
\caption{Left two panels: The SFH and AMR derived for \lp\ based on the F090W-F150W CMD (blue) and the F090W-F277W CMD (orange) using the PARSEC  (solid lines) and BaSTI stellar libraries (dashed lines). Shaded regions are shown for the PARSEC SFH solutions and represent statistical uncertainties only; statistical uncertainties on the BaSTI solutions have similar amplitudes.} The range in solutions between the two models gives an indication of the systematic uncertainties. The approximate epoch of reionization is marked with a vertical orange bar and the dotted diagonal line represents a constant star formation rate. The shaded regions on the AMR represent the resolution of 0.15 dex used in the fitting procedure. Right panel: The SFH based on the F090W-F150W catalog and the PARSEC library (blue solid) and BaSTI library (blue dashed) compared with the SFH derived from the shallower HST ACS F475W-F814W data and the PARSEC library (purple dotted). The shaded purple envelope represents combined statistical and systematic uncertainties on the HST SFH solution. See Section~\ref{sec:sfh_results} for discussion.
\label{fig:sfh}
\end{figure*}

Figure~\ref{fig:sfh} presents the SFH (left) and AMR (middle) from both the PARSEC and BaSTI libraries (see legend for details). The shaded regions represent the statistical uncertainties on the PARSEC solution; to avoid overlapping lines in the figure, we do not plot the statistical uncertainties on the BaSTI SFH solution, but they are of similar amplitude to the ones shown for PARSEC. The vertical shaded orange bar marks the approximate epoch of reionization. The difference between the solutions gives an indication of the systematic uncertainties, but, as discussed in Section~\ref{sec:sfh_method}, this is likely an underestimate. The SFH solutions between the stellar libraries and the different filter combinations are in excellent agreement. 

Overall, the AMR solutions from both libraries for the F090W-F150W CMD (blue lines) are also in good agreement with each other. The main result, regardless of the filter combination and stellar library, is that there has been little chemical evolution in \lp. Given the agreement between the PARSEC and BaSTI solutions for both the SFH and the AMR, in our downstream analysis, we adopt as representative the solutions based on only the PARSEC library. The shaded regions on the AMRs represent the resolution of 0.15 dex used in the fitting procedure, which are much larger than the statistical uncertainties and indicate the precision of the AMR solution. The AMRs derived from the two libraries for the F090W-F277W CMD are also quite close to one another, but with a slightly lower ($\sim0.2$ dex) present-day metallicity value relative to the F090W-F150W results.

In the right-hand panel of Figure~\ref{fig:sfh}, we compare our SFH solution derived based on the F090W-F150W CMD from PARSEC (blue solid line) and BaSTI (blue dashed line) with that derived from shallower HST optical imaging in the F475W and F814W filters using the PARSEC library (purple dotted line) from \citet{McQuinn2015a}. The light-purple shaded region represents the systematic uncertainties of the HST-based SFH, which increase significantly at older lookback times as the CMD is a few magnitudes shy of the oMSTO. The JWST and HST based solutions are in generally good agreement, with the JWST solution preferring a slightly slower build up of stellar mass at intermediate ages (e.g., $\sim$ 5-9 Gyr ago).

\subsection{Revised Stellar Mass of \lp}\label{sec:stellarmass}
The stellar mass of Leo P can be estimated by integrating the SFH over time and assuming a recycling fraction. This approach was applied to the SFH results based on the HST imaging with a reported present-day stellar mass value of $5.6^{+0.4}_{-1.9}\times10^5$ \msun\ \citep{McQuinn2015a}. This previous work assumed a recycling fraction of 30\% \citep{Kennicutt1994} and IMF limits of 0 to $\infty$ set by {\sc match}. Here, we use the SFH results derived from the F090W-F150W filters and the PARSEC library, but adopt a higher recycling value of 43\% appropriate for a metal-poor population \citep{Vincenzo2016} and scale our results to a Kroupa IMF with mass limits of 0.1 to 100 \msun\ \citep[e.g.,][]{Telford2020}. The result is a lower present-day stellar mass in Leo P of $2.9^{+0.6}_{-0.3}\times10^5$ \msun, also reported in Table~\ref{tab:properties}, which is slightly below the value of $5.7^{+0.4}_{-0.9}\times10^5$ \msun\ (considering the uncertainties) estimated based on a simply optical mass-to-light methodology \citep{McQuinn2013}. We find values comparable to our PARSEC-based stellar mass estimates using the results based on the BaSTI library and from the F090W-F277W filter combination for both. models. Note that, if we made the same assumptions from \citet{McQuinn2015a}, the stellar mass based on the SFH using NIRCam data would be in agreement within the uncertainties with the result from the HST data. 

\section{Discussion}\label{sec:discuss}
\subsection{\lp: Three phases of star formation activity}
Figure~\ref{fig:sfh} presents the first SFH of any galaxy outside the Local Group derived from a CMD that reaches below the oMSTO. This SFH provides unique insights into the mass assembly of an isolated low-mass galaxy that is outside a group environment and is not a satellite of a massive galaxy. 

Qualitatively, the SFH of \lp\ can be characterized by three phases: (i) an early onset of star formation; (ii) an extended pause after the reionization era; (iii) a re-ignition in star formation with a marked increase in activity that continues to the present-day.  As noted in Section \ref{sec:cmd}, signatures of these three main phases in the recovered SFH are seen in the distribution of stars in the CMD in Figure~\ref{fig:cmds_isochrones}. In particular, the sub-giant branch shows a change in density: there is a higher density of sources coincident with the isochrone of $\sim7$ Gyr and a lower density of stars towards the placement of the older-age isochrone. In addition, the presence of upper main sequence stars as well as blue and red helium burning stars are unambiguous signs of recent star formation activity over the last few 100 Myr. 

Quantitatively, we measure the durations of these phases by identifying the times of greatest change in the SFH. We report the values from the fits to the F090W-F150W CMD using the PARSEC stellar library but note that the phases are seen in the fits using the F090W$-$F277W data as well as the BaSTI library with the different filter combinations and with similar durations. 

Specifically, an early star formation epoch is identified across our first time bin, corresponding to a lookback time of $\sim12.6$ Gyr ($z\sim5$) when $15^{+0.1}_{-0.3}$ \% of the stellar mass formed. Note that, at this lookback time, our temporal resolution is limited. Therefore, we are unable to determine whether this stellar mass was formed, for example, all at once, in repeated short bursts, or in a more continuous fashion across this time period. \lp\ then experienced a lull, or pause in star formation for $\sim2.5$ Gyr (lookback times from $\sim12.5-10$ Gyr; $z\sim5-1.8$) before vigorously re-igniting. For a more direct comparison with other galaxies in the literature, we also calculate the lookback time at which \lp\ formed specific fractions of its stellar mass, namely 10\%, 25\%, 50\%, and 80\%. These timescales are often referred to as `$\tau _{\#\%}$' and help create a standardized metric to compare the SFHs of galaxies that can have a range of properties (e.g., different masses or different environments). We provide these values for the SFH fit from the F090W-F150W data with the PARSEC models in Table~\ref{tab:properties}; values from the BaSTI fits and from the F090W-F277W filters are comparable.

\begin{figure}
\begin{center}
\includegraphics[width=0.45\textwidth]{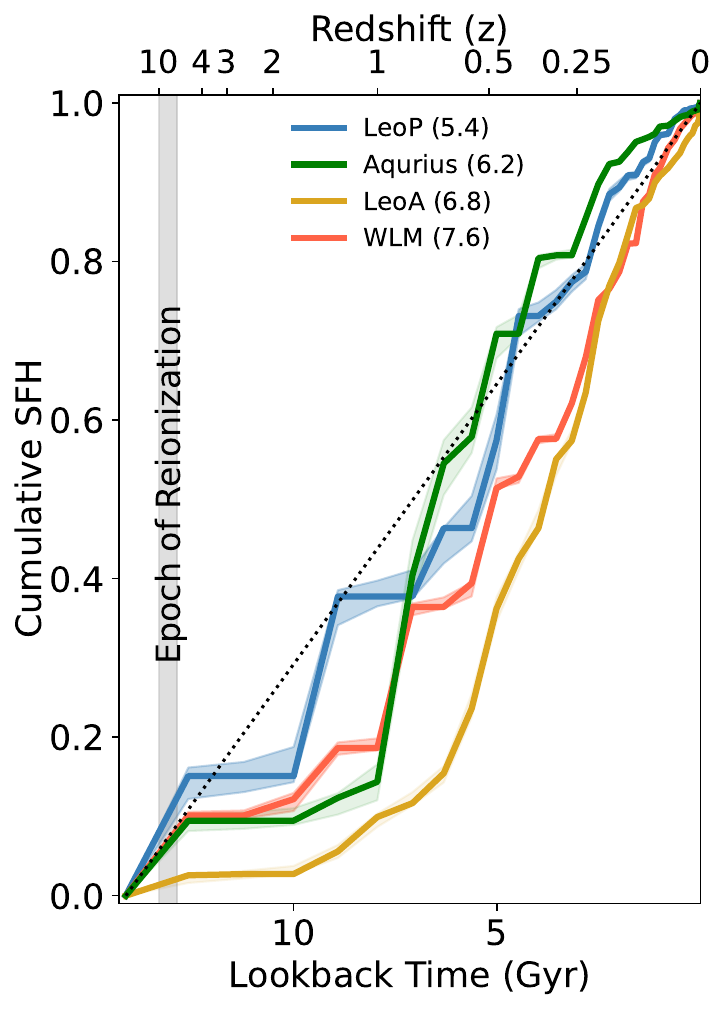}
\end{center}
\caption{The SFH of \lp\ based on the F090W-F150W CMD and the PARSEC library compared with the SFHs of the gas-rich isolated low-mass galaxies Aquarius, Leo~A, and WLM. The SFHs for Aquarius and Leo~A were derived from deep HST imaging \citep{Cole2014, Cole2007} and the SFH for WLM was derived from deep JWST NIRCam imaging \citep{McQuinn2024}. The shaded regions for each solution represent only the statistical uncertainties. The log of the present-day stellar mass in units of \msun\ are listed next to the galaxy names. The approximate epoch of reionization is shown as a shaded grey bar and the dotted line represents a constant star formation rate across all lookback times.}
\label{fig:sfh_compare}
\end{figure}

\begin{table*}
\begin{center}
\caption{Cumulative Star Formation Histories}
\label{tab:sfh}
\end{center}
\begin{center}
\vspace{-15pt}
\hspace{-1in}
\begin{tabular}{ll | c |c | c |c}   
\hline
\hline
log(t$_i$) & log(t$_f$) & Leo~P & Aquarius  & Leo~A  & WLM \\
\hline
\hline
6.6 & 6.7 & 1.0  & 1.0  & 1.0  & 1.0 \\
6.7 & 6.8 & 0.9994 $^{+ 0.0006 }_{- 0.0 }$ & 0.9987 $^{+ 0.0011 }_{- 0.0 }$ $^{+ 0.0012 }_{- 0.001 }$ & 0.9909 $^{+ 0.0012 }_{- 0.0 }$ $^{+ 0.0041 }_{- 0.0 }$ & 0.9962 $^{+ 0.0002 }_{- 0.0 }$ \\
6.8 & 6.9 & 0.9994 $^{+ 0.0006 }_{- 0.0001 }$ & 0.9987 $^{+ 0.0009 }_{- 0.0 }$ $^{+ 0.001 }_{- 0.001 }$ & 0.9909 $^{+ 0.0009 }_{- 0.0003 }$ $^{+ 0.004 }_{- 0.0003 }$ & 0.9962 $^{+ 0.0002 }_{- 0.0 }$ \\
6.9 & 7.0 & 0.9994 $^{+ 0.0006 }_{- 0.0002 }$ & 0.9987 $^{+ 0.0008 }_{- 0.0 }$ $^{+ 0.0009 }_{- 0.001 }$ & 0.9909 $^{+ 0.0007 }_{- 0.0005 }$ $^{+ 0.004 }_{- 0.0005 }$ & 0.9962 $^{+ 0.0001 }_{- 0.0001 }$ \\
7.0 & 7.1 & 0.9994 $^{+ 0.0004 }_{- 0.0005 }$ & 0.9987 $^{+ 0.0007 }_{- 0.0001 }$ $^{+ 0.001 }_{- 0.0005 }$ & 0.9909 $^{+ 0.0004 }_{- 0.0008 }$ $^{+ 0.004 }_{- 0.0008 }$ & 0.9962 $^{+ 0.0001 }_{- 0.0001 }$ \\
7.1 & 7.2 & 0.9994 $^{+ 0.0002 }_{- 0.0008 }$ & 0.9987 $^{+ 0.0005 }_{- 0.0003 }$ $^{+ 0.0007 }_{- 0.0006 }$ & 0.9909 $^{+ 0.0001 }_{- 0.0011 }$ $^{+ 0.004 }_{- 0.0011 }$ & 0.9962 $^{+ 0.0001 }_{- 0.0003 }$ \\
7.2 & 7.3 & 0.9994 $^{+ 0.0001 }_{- 0.0011 }$ & 0.9987 $^{+ 0.0003 }_{- 0.0005 }$ $^{+ 0.0005 }_{- 0.0007 }$ & 0.9909 $^{+ 0.0 }_{- 0.0014 }$ $^{+ 0.0038 }_{- 0.0014 }$ & 0.9955 $^{+ 0.0005 }_{- 0.0004 }$ \\
7.3 & 7.4 & 0.9994 $^{+ 0.0 }_{- 0.0015 }$ & 0.9984 $^{+ 0.0004 }_{- 0.0006 }$ $^{+ 0.0006 }_{- 0.0009 }$ & 0.9909 $^{+ 0.0 }_{- 0.0017 }$ $^{+ 0.004 }_{- 0.0017 }$ & 0.9942 $^{+ 0.0005 }_{- 0.0007 }$ \\
7.4 & 7.5 & 0.9987 $^{+ 0.0001 }_{- 0.0016 }$ & 0.9984 $^{+ 0.0002 }_{- 0.001 }$ $^{+ 0.0003 }_{- 0.0014 }$ & 0.9909 $^{+ 0.0 }_{- 0.0023 }$ $^{+ 0.0036 }_{- 0.0023 }$ & 0.9931 $^{+ 0.0007 }_{- 0.0003 }$ \\
7.5 & 7.6 & 0.9987 $^{+ 0.0 }_{- 0.0022 }$ & 0.9977 $^{+ 0.0005 }_{- 0.0013 }$ $^{+ 0.0011 }_{- 0.0014 }$ & 0.9909 $^{+ 0.0 }_{- 0.0037 }$ $^{+ 0.0036 }_{- 0.0037 }$ & 0.9931 $^{+ 0.0004 }_{- 0.0006 }$ \\
7.6 & 7.7 & 0.9945 $^{+ 0.0024 }_{- 0.0 }$ & 0.9918 $^{+ 0.0007 }_{- 0.0005 }$ $^{+ 0.0028 }_{- 0.0021 }$ & 0.9776 $^{+ 0.0038 }_{- 0.0 }$ $^{+ 0.0042 }_{- 0.0024 }$ & 0.9909 $^{+ 0.0005 }_{- 0.0005 }$ \\
7.7 & 7.8 & 0.9945 $^{+ 0.0015 }_{- 0.0005 }$ & 0.9918 $^{+ 0.0004 }_{- 0.0006 }$ $^{+ 0.0027 }_{- 0.0023 }$ & 0.9776 $^{+ 0.0007 }_{- 0.0019 }$ $^{+ 0.002 }_{- 0.0033 }$ & 0.9898 $^{+ 0.0003 }_{- 0.0003 }$ \\
7.8 & 7.9 & 0.9945 $^{+ 0.0012 }_{- 0.0008 }$ & 0.9915 $^{+ 0.0006 }_{- 0.0004 }$ $^{+ 0.0031 }_{- 0.0019 }$ & 0.9751 $^{+ 0.0008 }_{- 0.0016 }$ $^{+ 0.0034 }_{- 0.0021 }$ & 0.9897 $^{+ 0.0003 }_{- 0.0003 }$ \\
7.9 & 8.0 & 0.9945 $^{+ 0.0009 }_{- 0.001 }$ & 0.9915 $^{+ 0.0005 }_{- 0.0005 }$ $^{+ 0.0041 }_{- 0.0009 }$ & 0.9739 $^{+ 0.0004 }_{- 0.0018 }$ $^{+ 0.0043 }_{- 0.0019 }$ & 0.9891 $^{+ 0.0002 }_{- 0.0004 }$ \\
8.0 & 8.1 & 0.9945 $^{+ 0.0008 }_{- 0.0011 }$ & 0.9913 $^{+ 0.0005 }_{- 0.0005 }$ $^{+ 0.0031 }_{- 0.0022 }$ & 0.9727 $^{+ 0.0005 }_{- 0.0017 }$ $^{+ 0.0049 }_{- 0.0017 }$ & 0.9887 $^{+ 0.0003 }_{- 0.0003 }$ \\
8.1 & 8.2 & 0.9945 $^{+ 0.0007 }_{- 0.0013 }$ & 0.9903 $^{+ 0.0006 }_{- 0.0004 }$ $^{+ 0.0019 }_{- 0.0014 }$ & 0.9717 $^{+ 0.0003 }_{- 0.0019 }$ $^{+ 0.0041 }_{- 0.002 }$ & 0.9878 $^{+ 0.0003 }_{- 0.0003 }$ \\
8.2 & 8.3 & 0.9934 $^{+ 0.0009 }_{- 0.0009 }$ & 0.9901 $^{+ 0.0005 }_{- 0.0005 }$ $^{+ 0.0019 }_{- 0.0013 }$ & 0.9668 $^{+ 0.001 }_{- 0.0014 }$ $^{+ 0.0068 }_{- 0.0014 }$ & 0.9863 $^{+ 0.0002 }_{- 0.0004 }$ \\
8.3 & 8.4 & 0.9934 $^{+ 0.0006 }_{- 0.0012 }$ & 0.9893 $^{+ 0.0005 }_{- 0.0005 }$ $^{+ 0.0012 }_{- 0.0021 }$ & 0.9616 $^{+ 0.0007 }_{- 0.0015 }$ $^{+ 0.0094 }_{- 0.0015 }$ & 0.9851 $^{+ 0.0003 }_{- 0.0003 }$ \\
8.4 & 8.5 & 0.9931 $^{+ 0.0006 }_{- 0.0012 }$ & 0.9873 $^{+ 0.0006 }_{- 0.0006 }$ $^{+ 0.0011 }_{- 0.003 }$ & 0.9593 $^{+ 0.0007 }_{- 0.0015 }$ $^{+ 0.0082 }_{- 0.0015 }$ & 0.9826 $^{+ 0.0003 }_{- 0.0003 }$ \\
8.5 & 8.6 & 0.9909 $^{+ 0.0008 }_{- 0.0012 }$ & 0.9848 $^{+ 0.0005 }_{- 0.0006 }$ $^{+ 0.0009 }_{- 0.0012 }$ & 0.9549 $^{+ 0.0007 }_{- 0.0015 }$ $^{+ 0.0082 }_{- 0.0015 }$ & 0.9784 $^{+ 0.0004 }_{- 0.0004 }$ \\
8.6 & 8.7 & 0.9909 $^{+ 0.0005 }_{- 0.0015 }$ & 0.9841 $^{+ 0.0006 }_{- 0.0006 }$ $^{+ 0.0006 }_{- 0.0021 }$ & 0.9483 $^{+ 0.0008 }_{- 0.0015 }$ $^{+ 0.0079 }_{- 0.0015 }$ & 0.9737 $^{+ 0.0003 }_{- 0.0005 }$ \\
8.7 & 8.8 & 0.985 $^{+ 0.0012 }_{- 0.0018 }$ & 0.9822 $^{+ 0.0006 }_{- 0.0007 }$ $^{+ 0.0007 }_{- 0.003 }$ & 0.9368 $^{+ 0.001 }_{- 0.0014 }$ $^{+ 0.0112 }_{- 0.0014 }$ & 0.9676 $^{+ 0.0005 }_{- 0.0005 }$ \\
8.8 & 8.9 & 0.9801 $^{+ 0.0017 }_{- 0.0021 }$ & 0.9775 $^{+ 0.0008 }_{- 0.0008 }$ $^{+ 0.0028 }_{- 0.0057 }$ & 0.9289 $^{+ 0.0012 }_{- 0.0016 }$ $^{+ 0.0095 }_{- 0.0016 }$ & 0.9537 $^{+ 0.0006 }_{- 0.0008 }$ \\
8.9 & 9.0 & 0.961 $^{+ 0.0026 }_{- 0.0021 }$ & 0.9715 $^{+ 0.0009 }_{- 0.0009 }$ $^{+ 0.0026 }_{- 0.0046 }$ & 0.918 $^{+ 0.0014 }_{- 0.002 }$ $^{+ 0.0077 }_{- 0.0038 }$ & 0.9431 $^{+ 0.0007 }_{- 0.0007 }$ \\
9.0 & 9.05 & 0.9595 $^{+ 0.0019 }_{- 0.0033 }$ & 0.9703 $^{+ 0.0011 }_{- 0.001 }$ $^{+ 0.0014 }_{- 0.0071 }$ & 0.9079 $^{+ 0.0024 }_{- 0.0016 }$ $^{+ 0.0087 }_{- 0.0059 }$ & 0.9176 $^{+ 0.001 }_{- 0.001 }$ \\
9.05 & 9.1 & 0.9506 $^{+ 0.0048 }_{- 0.0036 }$ & 0.9611 $^{+ 0.0013 }_{- 0.0021 }$ $^{+ 0.0098 }_{- 0.004 }$ & 0.8996 $^{+ 0.0024 }_{- 0.0026 }$ $^{+ 0.0068 }_{- 0.0082 }$ & 0.9098 $^{+ 0.0011 }_{- 0.0011 }$ \\
9.1 & 9.15 & 0.9299 $^{+ 0.004 }_{- 0.0042 }$ & 0.9573 $^{+ 0.0016 }_{- 0.0014 }$ $^{+ 0.0033 }_{- 0.0083 }$ & 0.8797 $^{+ 0.003 }_{- 0.0018 }$ $^{+ 0.0052 }_{- 0.0174 }$ & 0.8847 $^{+ 0.0011 }_{- 0.0015 }$ \\
9.15 & 9.2 & 0.9254 $^{+ 0.0037 }_{- 0.0047 }$ & 0.9542 $^{+ 0.0017 }_{- 0.0015 }$ $^{+ 0.0017 }_{- 0.014 }$ & 0.8712 $^{+ 0.0027 }_{- 0.0021 }$ $^{+ 0.0055 }_{- 0.0219 }$ & 0.8756 $^{+ 0.0011 }_{- 0.0011 }$ \\
9.2 & 9.25 & 0.9089 $^{+ 0.0047 }_{- 0.0038 }$ & 0.951 $^{+ 0.0024 }_{- 0.0017 }$ $^{+ 0.0083 }_{- 0.0083 }$ & 0.8676 $^{+ 0.0011 }_{- 0.0046 }$ $^{+ 0.0012 }_{- 0.0275 }$ & 0.8231 $^{+ 0.0016 }_{- 0.0009 }$ \\
9.25 & 9.3 & 0.9089 $^{+ 0.0027 }_{- 0.0061 }$ & 0.9386 $^{+ 0.0025 }_{- 0.0031 }$ $^{+ 0.0043 }_{- 0.0219 }$ & 0.8343 $^{+ 0.0069 }_{- 0.0026 }$ $^{+ 0.0069 }_{- 0.036 }$ & 0.8226 $^{+ 0.001 }_{- 0.0018 }$ \\
9.3 & 9.35 & 0.8935 $^{+ 0.0089 }_{- 0.0049 }$ & 0.9259 $^{+ 0.0026 }_{- 0.0027 }$ $^{+ 0.0109 }_{- 0.0172 }$ & 0.7983 $^{+ 0.0043 }_{- 0.0057 }$ $^{+ 0.0043 }_{- 0.0314 }$ & 0.787 $^{+ 0.0025 }_{- 0.0019 }$ \\
9.35 & 9.4 & 0.885 $^{+ 0.0065 }_{- 0.0086 }$ & 0.9231 $^{+ 0.0022 }_{- 0.0032 }$ $^{+ 0.004 }_{- 0.0342 }$ & 0.7696 $^{+ 0.0062 }_{- 0.0046 }$ $^{+ 0.0062 }_{- 0.0368 }$ & 0.7656 $^{+ 0.0019 }_{- 0.0027 }$ \\
9.4 & 9.45 & 0.8443 $^{+ 0.0094 }_{- 0.0089 }$ & 0.8974 $^{+ 0.0042 }_{- 0.0033 }$ $^{+ 0.0089 }_{- 0.036 }$ & 0.725 $^{+ 0.0067 }_{- 0.0043 }$ $^{+ 0.0067 }_{- 0.0457 }$ & 0.7513 $^{+ 0.0026 }_{- 0.0021 }$ \\
9.45 & 9.5 & 0.7871 $^{+ 0.01 }_{- 0.0094 }$ & 0.8547 $^{+ 0.0038 }_{- 0.0056 }$ $^{+ 0.015 }_{- 0.0499 }$ & 0.635 $^{+ 0.0068 }_{- 0.0063 }$ $^{+ 0.0068 }_{- 0.0485 }$ & 0.6795 $^{+ 0.0026 }_{- 0.0028 }$ \\
9.5 & 9.55 & 0.7741 $^{+ 0.0088 }_{- 0.0122 }$ & 0.8078 $^{+ 0.0072 }_{- 0.0026 }$ $^{+ 0.0488 }_{- 0.0477 }$ & 0.5739 $^{+ 0.0132 }_{- 0.0028 }$ $^{+ 0.0191 }_{- 0.0262 }$ & 0.6211 $^{+ 0.0021 }_{- 0.0052 }$ \\
9.55 & 9.6 & 0.7493 $^{+ 0.0148 }_{- 0.0095 }$ & 0.8078 $^{+ 0.004 }_{- 0.0055 }$ $^{+ 0.0216 }_{- 0.0712 }$ & 0.5505 $^{+ 0.0012 }_{- 0.0206 }$ $^{+ 0.0067 }_{- 0.0404 }$ & 0.5762 $^{+ 0.0059 }_{- 0.0016 }$ \\
9.6 & 9.65 & 0.7312 $^{+ 0.0171 }_{- 0.0069 }$ & 0.8043 $^{+ 0.0038 }_{- 0.0129 }$ $^{+ 0.0683 }_{- 0.0613 }$ & 0.4638 $^{+ 0.023 }_{- 0.0 }$ $^{+ 0.0405 }_{- 0.013 }$ & 0.5762 $^{+ 0.0026 }_{- 0.0056 }$ \\
9.65 & 9.7 & 0.7312 $^{+ 0.0088 }_{- 0.0252 }$ & 0.709 $^{+ 0.0239 }_{- 0.0044 }$ $^{+ 0.0877 }_{- 0.0912 }$ & 0.4253 $^{+ 0.0079 }_{- 0.0158 }$ $^{+ 0.0186 }_{- 0.0274 }$ & 0.5286 $^{+ 0.003 }_{- 0.0079 }$ \\
9.7 & 9.75 & 0.5746 $^{+ 0.0324 }_{- 0.0371 }$ & 0.709 $^{+ 0.0083 }_{- 0.0311 }$ $^{+ 0.0198 }_{- 0.1445 }$ & 0.3622 $^{+ 0.0148 }_{- 0.0165 }$ $^{+ 0.0268 }_{- 0.0339 }$ & 0.5141 $^{+ 0.0123 }_{- 0.0002 }$ \\
9.75 & 9.8 & 0.4637 $^{+ 0.0404 }_{- 0.0167 }$ & 0.5789 $^{+ 0.0371 }_{- 0.0205 }$ $^{+ 0.2234 }_{- 0.0462 }$ & 0.2362 $^{+ 0.0184 }_{- 0.0113 }$ $^{+ 0.0651 }_{- 0.0114 }$ & 0.3941 $^{+ 0.0013 }_{- 0.0166 }$ \\
9.8 & 9.85 & 0.4637 $^{+ 0.0005 }_{- 0.0446 }$ & 0.5453 $^{+ 0.0295 }_{- 0.0399 }$ $^{+ 0.0661 }_{- 0.3222 }$ & 0.1545 $^{+ 0.009 }_{- 0.0118 }$ $^{+ 0.0926 }_{- 0.0119 }$ & 0.3643 $^{+ 0.012 }_{- 0.0019 }$ \\
9.85 & 9.9 & 0.3774 $^{+ 0.0334 }_{- 0.0024 }$ & 0.4045 $^{+ 0.0437 }_{- 0.0298 }$ $^{+ 0.0829 }_{- 0.2469 }$ & 0.1167 $^{+ 0.0139 }_{- 0.0029 }$ $^{+ 0.0972 }_{- 0.003 }$ & 0.3643 $^{+ 0.0043 }_{- 0.0106 }$ \\
9.9 & 9.95 & 0.3774 $^{+ 0.02 }_{- 0.0124 }$ & 0.1436 $^{+ 0.0222 }_{- 0.0233 }$ $^{+ 0.1774 }_{- 0.1193 }$ & 0.0992 $^{+ 0.0044 }_{- 0.0126 }$ $^{+ 0.0826 }_{- 0.0133 }$ & 0.1864 $^{+ 0.012 }_{- 0.003 }$ \\
9.95 & 10.0 & 0.3774 $^{+ 0.008 }_{- 0.0362 }$ & 0.1234 $^{+ 0.0059 }_{- 0.0211 }$ $^{+ 0.0468 }_{- 0.1153 }$ & 0.0558 $^{+ 0.0082 }_{- 0.0083 }$ $^{+ 0.0869 }_{- 0.0083 }$ & 0.1864 $^{+ 0.0069 }_{- 0.0087 }$ \\
10.0 & 10.05 & 0.1509 $^{+ 0.0367 }_{- 0.0077 }$ & 0.0946 $^{+ 0.0157 }_{- 0.0054 }$ $^{+ 0.0456 }_{- 0.0855 }$ & 0.0273 $^{+ 0.0102 }_{- 0.0 }$ $^{+ 0.0955 }_{- 0.0001 }$ & 0.1218 $^{+ 0.0078 }_{- 0.0151 }$ \\
10.05 & 10.1 & 0.1509 $^{+ 0.0178 }_{- 0.0202 }$ & 0.0946 $^{+ 0.0104 }_{- 0.0103 }$ $^{+ 0.0281 }_{- 0.0738 }$ & 0.0272 $^{+ 0.0041 }_{- 0.0054 }$ $^{+ 0.082 }_{- 0.0054 }$ & 0.1012 $^{+ 0.0063 }_{- 0.004 }$ \\
10.1 & 10.15 & 0.1509 $^{+ 0.0107 }_{- 0.0289 }$ & 0.0946 $^{+ 0.0079 }_{- 0.0131 }$ $^{+ 0.0272 }_{- 0.0744 }$ & 0.0257 $^{+ 0.0 }_{- 0.0097 }$ $^{+ 0.0815 }_{- 0.0097 }$ & 0.1012 $^{+ 0.0045 }_{- 0.0058 }$ \\
\hline
\hline
\end{tabular} 
\end{center}
\tablecomments{Fraction of stellar formed as a function of time and the statistical uncertainties for the four low-mass galaxies \lp, Aquarius, Leo~A, and WLM. For Aquarius and Leo~A, whose SFHs were measured from HST data, we also provide the total uncertainties (i.e., statistical uncertainties combined in quadrature with systematic uncertainties estimated from Monte Carlo simulations; \citet[see][for details]{Dolphin2013}. A machine readable version of this table is available online.}
\end{table*}

\subsection{Comparison with other isolated gas-rich low-mass galaxies}
While \lp\ is the first galaxy outside the Local Group with a SFH from imaging reaching below the oMSTO, there are three gas-rich ($M_{\rm HI}$/\mstar\ $> $1), low-mass (\mstar\ $\ltsimeq 10^8$ \msun) galaxies within the Local Group that are considered isolated and that also have the requisite data needed for a robustly measured lifetime SFH.\footnote{Other star-forming galaxies in the Local Group include IC10, IC1613, NGC6822, and the Small and Large Magellanic Clouds. We do not consider these in our comparison as they are either more massive (\mstar $\gtsimeq 10^8$ \msun), known to be interacting with another system(s) (i.e., not isolated), or both.}  These galaxies are: Aquarius (\mstar\ $=1.6 \times10^6$ \msun), Leo~A (\mstar\ $=6.0 \times10^6$ \msun), and WLM (\mstar\ $=4.3 \times10^7$ \msun). All three of these galaxies are located in the outskirts of the Local Group and are considered isolated based on their locations relative to other known systems and their measured velocities \citep{McConnachie2012} and have a low probability of previous tidal interactions \citep[e.g.,][]{Shaya2013}.

Figure~\ref{fig:sfh_compare} compares the SFH of \lp\ with the SFHs of these three isolated gas-rich, low-mass galaxies. The SFHs for Leo~A and Aquarius are based on HST ACS imaging that cover the majority of the stellar disks of the galaxies \citep{Cole2007, Cole2014} but the solutions have been updated with the PARSEC stellar library and assume a Kroupa IMF which enables a more direct comparison with our \lp\ results. These updated SFHs are in close agreement with the previously published results. The SFH of WLM is based on JWST NIRCam data, again with the PARSEC library and the same Kroupa IMF \citep{McQuinn2024}. Note that the spatial coverage of the NIRCam imaging samples only the southern half of the main stellar disk of WLM, bringing into question whether the SFH from these data are representative of the full galaxy.  However, the SFH derived from ACS imaging of WLM that extends the spatial coverage out to larger radii was shown to have the same features - both qualitatively and quantitatively -  that we derive here \citep{Albers2019}. Furthermore, SFHs derived from imaging of off-axis regions of both the eastern (obtained with the HST WFC3 instrument taken in parallel with the ACS data) and western (obtained with the JWST NIRISS instrument taken in parallel with the NIRCAM data presented here) sides of WLM show phases quite similar to what we find in the main stellar disk, albeit with different amplitudes (Roger E. Cohen et al., in preparation). Thus, despite the incomplete spatial coverage of WLM, the existing results suggest the SFH of WLM does indeed follow the same pattern of early star formation, an extended pause, followed by reignition.

 Table~\ref{tab:sfh} provides the SFH solutions of the four systems with statistical uncertainties; we also provide the total uncertainties (i.e., combined statistical and systematic uncertainties) for Leo~A and Aquarius which were based on the HST data. For completeness, we also provide the AMRs in Table~\ref{tab:amr} in the Appendix. From Figure~\ref{fig:sfh_compare}, all three galaxies show a similar pattern in their SFHs as \lp, namely the three phases described above (early star formation, extended pause, vigorous re-ignition), although it is worth noting that Leo~A shows a very slow start to star formation relative to the other galaxies. Indeed, \citet{Cole2007} characterized the SFH of Leo~A as being delayed, which, given its weak activity at early times, is an apt description. 

There is one additional galaxy with similar SFH features that we ultimately exclude from our comparison. As mentioned in Section~\ref{sec:environment}, Leo~T is a gas-rich low-mass galaxy in the Local Group. While not isolated at the present-day, it is thought to be on its first infall to the Local Group. The SFH has been derived based on relatively deep WFCP2 data, and it shows a similar pattern as seen in our comparison sample \citep{Clementini2012, Weisz2012}. However, the uncertainties on the SFH are quite large, likely driven by the poorer quality of the WFPC2 imaging relative to the ACS or NIRCam data and slightly shallower depth. Thus, we do not include Leo~T in our detailed comparison, but note that the SFH results of Leo~T support the trends seen in Figure~\ref{fig:sfh_compare}. 

The fact that the SFHs in Figure~\ref{fig:sfh_compare} have similar patterns is somewhat surprising. The a priori expectation is that the mass assembly of low-mass galaxies is a stochastic process, which should result in a diversity of SFHs even in a sample of four. Leo~A and Aquarius have been assumed to be outliers in their `late-blooming' SFHs. Previous SFHs of a larger sample of nearby dwarf galaxies reported on the diversity of SFHs with differences noted within a given morphological type \citep{Weisz2011}. This work also noted that the isolated dwarf irregulars spanning the same mass range formed the majority of their stellar mass before $z\sim1$, which we do not find. It is possible that the four galaxies in Figure~\ref{fig:sfh_compare} are serendipitously similar and not representative of the larger population of isolated dwarfs. On the other hand, the previous results were derived from shallower data and the SFHs have large uncertainties, particularly at older lookback times. Thus, it's possible that the pattern in the SFH of low-mass galaxies seen in Figure~\ref{fig:sfh_compare} is more common than previously inferred.

\subsection{The Masses of the Galaxies at the Time of Reionization}
Discussions of the perceived boundary between low-mass galaxies which are totally quenched at the time of reionization (often associated with the label ``ultrafaint'') and more massive galaxies which continue to form stars often concentrate on the stellar masses of these galaxies \citep[e.g.,][]{Bullock2017,Simon2019}.  These reviews settled on a luminosity definition of ultrafaint galaxies of those fainter than M$_V$ $= -7.7$, which corresponds to a luminosity of 10$^5$ L$_\odot$ or a stellar mass of roughly 10$^5$ M$_\odot$. For totally quenched galaxies, these stellar masses are representative of the stars that formed up to reionization.  However, for galaxies which were not totally quenched, the present day stellar mass can be very different from that at the epoch of reionization.

With lifetime SFHs, we have the opportunity to investigate the stellar masses of the low-mass galaxies in the ``transition zone'' between totally quenched galaxies and more massive galaxies at the time of reionization and to compare them to the masses of ultrafaint galaxies.  The four galaxies in Figure~\ref{fig:sfh_compare} had stellar masses of 4.4 $\times$ 10$^4$ M$_\odot$ (Leo~P), 1.5 $\times$ 10$^5$ M$_\odot$ (Leo~A), 1.5 $\times$ 10$^5$ M$_\odot$ (Aquarius), and 4.3 $\times$ 10$^6$ M$_\odot$ (WLM) at the time of reionization based on integrating the best-fitting SFHs. Thus, at the epoch of reionization, \lp\ had a stellar mass which was below the mass boundary currently associated with complete quenching by reionization. Leo~A and Aquarius were on that boundary and WLM was a decade above it. 

This comparison reveals two rather remarkable insights.  First, it becomes clear that the stellar mass (which is a proxy for the halo mass) at the time of reionization is likely not the sole property determining whether a galaxy is to be totally quenched. It is possible that the role of environment (e.g., a galaxy being isolated versus being a satellite) results in a different boundary value \citep[e.g.,][]{Christensen2024}. Second, the range of stellar masses at the time of reionization which can lead to the observed pattern of a pause followed by re-ignition is relatively large (at least two orders of magnitude).  Many studies have noted the large range in SFH properties for dwarf galaxies \citep[e.g.,][]{Weisz2015, Albers2019}, but it appears that the SFHs for truly isolated galaxies in this mass transition zone may show a uniformity of SFH patterns.

Since the reionization of the universe proceeded in a highly inhomogeneous way from $z \sim 10 - 5.3$ \citep[e.g.,][]{Bosman2021, Bosman2022}, it may be too simplistic to expect a sharp defining separation between low-mass galaxies that quench during reionization and those that do not. Sometimes described as a `patchy' or `inside-out' process, the ionization fronts produced by galaxies in over-dense regions occur first and are the strongest. These ionization fronts move out slowly through gas with overdensities, and then sweep through lower density regions at higher speeds at a later time \citep[e.g.,][]{Kannan2022, Lu2024}. While our Local Group does not reside in a void, it is located in a relatively under-dense region, and the galaxies in question were likely impacted by reionization at a somewhat later timescale than galaxies in denser regions closer to the MW and M31. This would have allowed such galaxies additional time to build up their mass, their gravitational potential, and their ability to partially self-shield. 

\subsection{How Extended Pauses in Star Formation May Impact the Galaxy UV Luminosity Function}
From Figure~\ref{fig:sfh_compare}, these isolated, low-mass galaxies show little star formation activity between $z\sim5-1$. This low level of star formation implies a correspondingly low production of UV photons during this epoch. In addition, nearly all galaxies studied to date with lower masses (log(\mstar/\msun) $ <5$) were quenched at early times and therefore also had a similarly low production of UV photons for $z\ltsimeq5$. 

This has very important implications for the evolution of the galaxy UV luminosity function and the location of the turn-down in the luminosity function at intermediate redshifts. If these local constraints are representative of the mass assembly of low-mass (present-day log(\mstar/\msun) $<8$) galaxies, this implies that only more massive galaxies contribute significantly to the galaxy UV luminosity function between $z\sim5-1$. The overall dearth of UV photons from low-mass galaxies would mean the number density of galaxies at lower UV luminosity will not increase substantially, which has been noted observationally in studies from $z\sim3-5$, alongside an increase in number density at higher luminosities \citep{Finkelstein2022}. 

To date, most of the focus on the contribution of low-mass galaxies to the ionizing continuum has been focused on their contribution to the reionization of the universe. A constraint from nearby galaxies that their UV photon contribution has decreased at intermediate redshifts is complementary to the work that has shown how Local Group galaxies were likely important contributors to reionizing the universe at the earlier epoch of $z\sim7$ \citep[e.g.,][]{BoylanKolchin2015}. It would also imply that the turn-down in the galaxy UV luminosity function at redshifts between $5>z>1$ would occur at brighter magnitudes (i.e., for higher galaxy masses) that tentatively seen at redshifts $z>6$ \citep{Atek2023}. The turn-down at brighter magnitudes would then be ameliorated at $z\sim<1$ as star formation re-ignites in galaxies with log(\mstar/\msun) $\gtsimeq 5$. Additional observations at the faint end of the UV luminosity function at intermediate redshifts are needed to confirm whether this bears out based on larger low-mass galaxy population studies.

\subsection{Star formation re-ignition mechanisms}
In all four galaxies in Figure~\ref{fig:sfh_compare}, the extended pause is followed by a significant increase in star formation. This re-ignition could have been triggered by environmental factors or be secular in nature. 

Focusing first just on \lp, the re-ignition of star formation occurred at a lookback time of $\sim10$ Gyr and again at $\sim7$ Gyr, the latter of which is similar to the estimated timing ($\sim7$ Gyr ago) of a speculated fly-by interaction with the Local Group (see Section~\ref{sec:environment}). In this scenario, a fly-by with the Local Group would naturally explain the marked increase in star formation in \lp. Our team is actively working on combining these JWST data with the existing HST ACS data to measure the proper motion of \lp\ and reconstruct its past orbital history. Such analysis will provide insight into the likelihood that this interaction occurred and provide better constraints about the timing of the interaction. Additionally, detailed SFHs of other galaxies in the 14$+$12 association based on deep imaging would be extremely valuable. If all systems show an increase in star formation activity synchronized in time, it would provide additional corroborating evidence of a past interaction with the Local Group. 

On the other hand, \lp\ is not the only system to have star formation reignited $\sim$8 Gyr ago. Both WLM and Aquarius show re-ignition in the same epoch. Leo~A, which experienced the slowest start to star formation, took an additional $\sim$Gyr before star formation activity increased significantly, but with an overall pattern consistent with the other three systems. The similarity in the duration of the pause may reflect the timescale for cooling a metal-poor interstellar medium sufficiently for cloud collapse to occur on a larger scale thereby re-igniting star formation. In addition, star formation could have been fueled further by gas accretion from the surrounding region over a similar timescale. Cosmological simulations suggest that low-mass halos at distances $\sim1-5$ R$_{virial}$ from a massive galaxy and in an environment similar to that of the galaxies studied here are likely to accrete denser gas from the cosmic web which subsequently leads to late-time star formation activity \citep{Chun2020}. Similarly, a different model suggests late outbursts of star formation ($z<2$) in low-mass galaxies can occur when the intensity of the intergalactic UV background decreases, which then enables late-time accretion from the intergalactic medium \citep{Ricotti2009}.

Another possible trigger mechanism for re-igniting star formation is a merger with another system. For the mass regime studied here ($10^5 - 10^7$ \msun), detailed analysis in the FIRE cosmological simulations find that galaxies undergo an average of five galaxy mergers, and that the majority of mergers occur early ($z>3$) and have mass ratios less than 1:10 \citep{Fitts2018}. The large mass differentials in the mergers means that the present-day simulated dwarf galaxies have at least 70\% and as much as 90\% of their stellar mass formed in-situ. Similarly low merger rates have been reported across other simulations, and most major mergers also occur early at $z<2$ \citep[e.g.,][]{Wright2019, Ledinauskas2018, Gandhi2023}. The predicted low probability of mergers and the preferred occurrence at early times is in tension with the late epoch and approximately simultaneous re-ignition of star formation seen for the four galaxies studied here. Thus, it is unlikely that mergers are the trigger mechanism for the marked increase in star formation activity noted in Figure~\ref{fig:sfh_compare}.

From a theoretical perspective, there are a number of other proposed explanations for late star formation activity in isolated low-mass galaxies. High-resolution cosmological simulations have recovered `gappy' SFHs with an extended pause followed by re-ignition where the re-ignition of star formation activity is attributed to a collision of the galaxy with the cosmic web or tidal gas streams left over from galaxy mergers \citep{Wright2019}. In these simulations, $\sim$20\% of galaxies have gappy SFHs, and approximately half of those gappy systems have SFHs similar to those shown in Figure~\ref{fig:sfh_compare} and, notably, comparable gas factions at $z=0$. 

Interestingly, the simulated galaxies with gappy SFHs in \citet{Wright2019} also had lower halo masses at $z\sim3$ and correspondingly lower virial temperatures than galaxies with more continuous activity. The lower virial temperature would mean that reionization and stellar feedback had a greater impact on the star formation conditions. Along similar lines, it has been proposed that the dark matter halos that host galaxies with significant star formation activity at late times may have formed later than the average halo, particularly when isolated \citep[e.g.,][]{Christensen2024}. Semi-analytic modelling by \citet{Ledinauskas2018} that follows the accretion of baryonic matter, star formation, and stellar feedback in dark matter halos was used to analyze the likelihood of producing SFHs like Leo~A and Aquarius, where the bulk of stellar mass is formed after $z\sim1$. Based on models of 1927 different galaxies, they produce a SFH like Leo~A 0.8\% and like Aquarius 2.8\% of the time. The models suggest that the late build up of mass is due to the late formation of the dark matter halo that hosts the galaxies; however, the modelled galaxies had 20-30 times more gas than Leo~A and Aquarius. It is unclear how this impacts the modelled SFHs or overall interpretation. Cosmological simulations have also produced late-forming halos, but the SFHs in these systems are not consistent with the pattern shown in Figure~\ref{fig:sfh_compare} \citep{Benitez-Llambay2015}. 

In summary, while it is not yet possible to definitively determine the star-formation re-ignition mechanism, given the similar timescales it seems more likely the cause is inherent to the growth of these galaxies (cooling timescales, halo mass and growth rate) rather than environmentally driven (e.g., fly-by with the Local Group, mergers, or interaction with surrounding, low-density gas).

\section{Conclusions}\label{sec:conclusions}
We present imaging of the resolved stars in the nearby low-mass, extremely metal-poor galaxy \lp\ obtained with the JWST NIRCam instrument in the F090W, F150W, and F277W filters. The imaging reaches photometric depths below the oMSTO, enabling accurate characterization of the early SFH in a galaxy outside the Local Group for the first time.

We summarize our findings as follows:
\begin{itemize}
\item We tested the efficacy of simultaneous NIRCam SW and LW imaging for SFH fitting purposes compared with 2 SW filters that require separate, and thus additive, integration times. Based on our observing strategy, the photometry from all three filters reached below the oMSTO and the SFHs derived separately from the F090W-F150W filter combination and the F090W-F277W filter combination are in good agreement with each other (Figure~\ref{fig:sfh}). However, despite reaching the requisite depths in all filters, the LW F277W filter had significantly ($\sim2\times$) longer exposure time and the F277W photometry is still shallower than the F090W and F150W (see exposure times and completeness limits in Table~\ref{tab:properties} and Figure~\ref{fig:cmds}) and is more impacted by crowding (Figure~\ref{fig:zoom_filters}). Thus, given the decreased efficiency in reaching the needed photometric depth in the LW F277W filter, observing programs targeting SFHs from deep imaging of resolved stars will require less time using the 2 SW filters (i.e., F090W and F150W) compared with using the SW F090W filter with simultaneous imaging the LW F277W filter. 

\item The SFH of \lp\ can be characterized by three phases: (i) an early onset of star formation; (ii) an extended pause post-reionization; (iii) a re-ignition in star formation with a marked increase in activity that continues to the present-day. The SFHs of the only other isolated, low-mass ($5<$log(\mstar/\msun)$<8$) galaxies (Leo~A, Aquarius, WLM) derived from similar quality data show similar phases (Figure~\ref{fig:sfh_compare}). 

\item While the present-day stellar masses of all four galaxies are greater than the galaxy mass boundary associated with complete quenching by reionization, \lp\ had a mass that was a factor of 10 below that at $z>5$, which is also lower than the stellar mass of many of the quenched ultrafaint dwarfs in the Local Group; two galaxies (Leo~A and Aquarius) had \mstar\ at the boundary of the ultrafaint dwarfs ($\sim10^5$ \msun). This provides strong evidence that is it not just the mass of the galaxy at the time of reionization that determines whether it will be quenched: the environment of the system (i.e., whether it is isolated or a satellite) is an important factor. First, the timing when reionization occurs locally may impact the process. Second, galaxies that are isolated are more likely to later (re-)accrete gas and reignite star formation, whereas galaxies in denser environments are located in halos of hot gas which inhibit the accretion and cooling of gas needed to form new stars. 

\item The galaxy masses lie in the range $5<$ (log(\mstar/\msun) $<8$ which appears to form a transition zone where reionization impacts, but does not stop, the stellar mass growth of the galaxies. If the extended pause seen in the SFHs is indeed due to regulation by reionization, it creates a testable prediction, namely that one would expect a similar signature in the SFHs of other isolated dwarfs of similar mass \citep{Grebel2004}. Upcoming work on two additional isolated galaxies in the Local Group using JWST data will help in this regard (Sag DIG, JWST-GO-5255; Sextans~A, JWST-AR-6118).

\item If the trend in extended periods of quiescence at intermediate redshifts (i.e., $z\sim5-1$)  in low-mass galaxies bears out with a larger sample, the contribution of galaxies with $5<$ (log(\mstar/\msun) $<8$ to the galaxy UV luminosity function will decrease post-reionization. Coupled with the result that all galaxies with present-day masses below log(\mstar/\msun) $=5$ are quenched at early times, we would expect the turn-down in the UV luminosity function to occur at higher masses between $z\sim5-1$, followed by an up-turn when the galaxies with log(\mstar/\msun) $>5$ re-ignite their star formation. 

\item We identify several mechanisms that could account for the re-ignition of star formation in \lp\ as well as the other three galaxies including (i) a previous fly-by with the Local Group, (ii) the intrinsic physical process of metal-poor gas cooling and eventually collapsing to form stars, possibly with additional fueling by gas accretion from the cosmic web, (iii) interactions with low-density gas, and (iv) past mergers. Other models focusing on the late build-up of stellar mass in low-mass galaxies (vs.\ a `re-ignition') indicate that the dark matter halos that host such galaxies also form late. While some proposed explanations appear less feasible than others (e.g., synchronous mergers for all four isolated galaxies seems less likely than a scenario involving long cooling times for metal-poor gas), the main factors that drive the vigorous re-ignition of star formation and late build-up of stellar mass in these galaxies remains uncertain.
\end{itemize}

\acknowledgments
Based on observations with the NASA/ESA James Webb Space Telescope obtained from MAST at the Space Telescope Science Institute, which is operated by the Association of Universities for Research in Astronomy, Incorporated, under NASA contract NAS5-26555. Support for this work was provided by NASA through grant No.\ JWST-GO-1617 from the Space Telescope Science Institute under NASA contract NAS5-26555. This research has made use of NASA Astrophysics Data System Bibliographic Services, adstex\footnote{https://github.com/yymao/adstex}, and the arXiv preprint server. 

\facilities{James Webb Space Telescope}
\software{This research made use of {\tt DOLPHOT} \citep{Dolphin2000, Weisz2024}, {\tt MATCH} \citep{Dolphin2002, Dolphin2012, Dolphin2013},  and Astropy\footnote{http://www.astropy.org}, a community-developed core Python package for Astronomy \citep{astropy:1801.02634}.}\\

\appendix
Table~\ref{tab:amr} provides the age-metallicity relations for \lp\ and the other three galaxies studied (Aquarius, Leo~A, and WLM) that were derived as part of the CMD-fitting process using the PARSEC stellar libary. 

\begin{table*}
\begin{center}
\caption{Age-Metallicity Relations}
\label{tab:amr}
\end{center}
\begin{center}
\vspace{-15pt}
\hspace{-1in}
\begin{tabular}{ll | c |c | c |c}  
\hline
\hline
log(t$_i$) & log(t$_f$) & Leo~P & Aquarius  & Leo~A  & WLM \\
\hline
\hline
6.6 & 6.7 & -1.217$\pm0.0.75$ & -1.273 $^{+ 0.578 }_{-0}$ & -1.225 $^{+ 0.365 }_{- 0.004}$ & -0.603 \\
6.7 & 6.8 & \nodata & \nodata & \nodata & \nodata \\
6.8 & 6.9 & \nodata & \nodata & \nodata & \nodata \\
6.9 & 7.0 & \nodata & \nodata & \nodata & \nodata \\
7.0 & 7.1 & \nodata & \nodata & \nodata & -0.603 \\
7.1 & 7.2 & \nodata & \nodata & \nodata & -0.603 \\
7.2 & 7.3 & \nodata & -1.274 $^{+ 0.291 }_{- 0.145}$ & \nodata & -0.604 \\
7.3 & 7.4 & -1.217$\pm0.0.75$ & \nodata & \nodata & -0.604 \\
7.4 & 7.5 & \nodata & -1.274 $^{+ 0.291 }_{- 0.145}$ & \nodata & \nodata \\
7.5 & 7.6 & -1.217$\pm0.0.75$ & -1.275 $^{+ 0.001 }_{- 0.289}$ & -1.249 $^{+ 0.183 }_{- 0.215}$ & -0.605 \\
7.6 & 7.7 & \nodata & \nodata & \nodata & -0.605 \\
7.7 & 7.8 & \nodata & -1.276 $^{+ 0.292 }_{-0}$ & -1.269 $^{+ 0.584 }_{- 0.001}$ & -0.606 \\
7.8 & 7.9 & \nodata & \nodata & -1.232 $^{+ 0.408 }_{- 0.053}$ & -0.607 \\
7.9 & 8.0 & \nodata & -1.277 $^{+ 0.144 }_{- 0.289}$ & -1.271 $^{+ 0.73 }_{- 0.001}$ & -0.608 \\
8.0 & 8.1 & \nodata & -1.279 $^{+ 0.289 }_{- 0.288}$ & -1.272 $^{+ 0.667 }_{- 0.001}$ & -0.609 \\
8.1 & 8.2 & -1.217$\pm0.0.75$ & -1.28 $^{+ 0.58 }_{-0}$ & -1.234 $^{+ 0.444 }_{- 0.146}$ & -0.61 \\
8.2 & 8.3 & \nodata & -1.282 $^{+ 0.287 }_{- 0.287}$ & -1.248 $^{+ 0.346 }_{- 0.125}$ & -0.612 \\
8.3 & 8.4 & -1.217$\pm0.0.75$ & -1.284 $^{+ 0.286 }_{- 0.286}$ & -1.241 $^{+ 0.391 }_{- 0.075}$ & -0.615 \\
8.4 & 8.5 & -1.217$\pm0.0.75$ & -1.287 $^{+ 0.184 }_{- 0.285}$ & -1.221 $^{+ 0.45 }_{- 0.041}$ & -0.618 \\
8.5 & 8.6 & \nodata & -1.291 $^{+ 0.292 }_{- 0.283}$ & -1.267 $^{+ 0.359 }_{- 0.1}$ & -0.622 \\
8.6 & 8.7 & -1.217$\pm0.0.75$ & -1.296 $^{+ 0.141 }_{- 0.281}$ & -1.253 $^{+ 0.117 }_{- 0.241}$ & -0.627 \\
8.7 & 8.8 & -1.217$\pm0.0.75$ & -1.302 $^{+ 0.28 }_{- 0.278}$ & -1.278 $^{+ 0.446 }_{- 0.089}$ & -0.634 \\
8.8 & 8.9 & -1.217$\pm0.0.75$ & -1.309 $^{+ 0.573 }_{-0}$ & -1.255 $^{+ 0.445 }_{- 0.01}$ & -0.642 \\
8.9 & 9.0 & -1.217$\pm0.0.75$ & -1.319 $^{+ 0.457 }_{-0}$ & -1.271 $^{+ 0.387 }_{- 0.02}$ & -0.652 \\
9.0 & 9.05 & -1.217$\pm0.0.75$ & -1.327 $^{+ 0.066 }_{- 0.269}$ & -1.306 $^{+ 0.194 }_{- 0.222}$ & -0.661 \\
9.05 & 9.1 & -1.217$\pm0.0.75$ & -1.334 $^{+ 0.197 }_{- 0.266}$ & -1.274 $^{+ 0.348 }_{- 0.094}$ & -0.668 \\
9.1 & 9.15 & -1.217$\pm0.0.75$ & -1.341 $^{+ }_{- 0.263}$ & -1.311 $^{+ 0.369 }_{- 0.111}$ & -0.676 \\
9.15 & 9.2 & -1.217$\pm0.0.75$ & -1.35 $^{+ 0.298 }_{- 0.259}$ & -1.26 $^{+ 0.328 }_{- 0.166}$ & -0.685 \\
9.2 & 9.25 & \nodata & -1.359 $^{+ 0.26 }_{- 0.255}$ & -1.333 $^{+ 0.414 }_{- 0.015}$ & -0.696 \\
9.25 & 9.3 & -1.217$\pm0.0.75$ & -1.37 $^{+ 0.246 }_{- 0.251}$ & -1.306 $^{+ 0.249 }_{- 0.127}$ & -0.707 \\
9.3 & 9.35 & -1.217$\pm0.0.75$ & -1.382 $^{+ 0.228 }_{- 0.246}$ & -1.349 $^{+ 0.337 }_{- 0.055}$ & -0.72 \\
9.35 & 9.4 & -1.217$\pm0.0.75$ & -1.395 $^{+ 0.13 }_{- 0.241}$ & -1.345 $^{+ 0.341 }_{- 0.012}$ & -0.735 \\
9.4 & 9.45 & -1.217$\pm0.0.75$ & -1.41 $^{+ 0.232 }_{- 0.235}$ & -1.361 $^{+ 0.325 }_{- 0.005}$ & -0.751 \\
9.45 & 9.5 & -1.217$\pm0.0.75$ & -1.427 $^{+ 0.139 }_{- 0.227}$ & -1.379 $^{+ 0.195 }_{- 0.087}$ & -0.769 \\
9.5 & 9.55 & -1.217$\pm0.0.75$ & \nodata & -1.411 $^{+ 0.388 }_{- 0.059}$ & -0.79 \\
9.55 & 9.6 & -1.217$\pm0.0.75$ & -1.467 $^{+ 0.107 }_{- 0.21}$ & -1.419 $^{+ 0.167 }_{- 0.115}$ & \nodata \\
9.6 & 9.65 & \nodata & -1.49 $^{+ 0.101 }_{- 0.201}$ & -1.442 $^{+ 0.272 }_{- 0.059}$ & -0.839 \\
9.65 & 9.7 & -1.217$\pm0.0.75$ & \nodata & -1.453 $^{+ 0.281 }_{- 0.03}$ & -0.868 \\
9.7 & 9.75 & -1.217$\pm0.0.75$ & -1.546 $^{+ 0.278 }_{- 0.176}$ & -1.495 $^{+ 0.223 }_{- 0.024}$ & -0.901 \\
9.75 & 9.8 & \nodata & -1.58 $^{+ 0.084 }_{- 0.161}$ & -1.529 $^{+ 0.182 }_{- 0.039}$ & -0.938 \\
9.8 & 9.85 & -1.217$\pm0.0.75$ & -1.617 $^{+ 0.152 }_{- 0.144}$ & -1.541 $^{+ 0.234 }_{- 0.061}$ & \nodata \\
9.85 & 9.9 & \nodata & -1.658 $^{+ 0.119 }_{- 0.123}$ & -1.655 $^{+ 0.409 }_{-0}$ & -1.025 \\
9.9 & 9.95 & \nodata & -1.703 $^{+ 0.083 }_{- 0.101}$ & -1.64 $^{+ 0.277 }_{- 0.01}$ & \nodata \\
9.95 & 10.0 & -1.217$\pm0.0.75$ & -1.751 $^{+ 0.478 }_{- 0.076}$ & -1.678 $^{+ 0.239 }_{- 0.123}$ & -1.136 \\
10.0 & 10.05 & \nodata & \nodata & -1.709 $^{+ 0.352 }_{- 0.057}$ & -1.201 \\
10.05 & 10.1 & \nodata & \nodata & -1.845 $^{+ 1.045 }_{-0}$ & \nodata \\
10.1 & 10.15 & -1.217$\pm0.0.75$ & -1.884 $^{+ 0.228 }_{- 0.007}$ & -1.824 $^{+ 0.389 }_{- 0.009}$ & -1.357 \\
\hline
\hline
\end{tabular} 
\end{center}
\vspace{-8pt}
\tablecomments{Age-Metallicity ([M/H]) Relations based on the best-fitting solutions to the CMDs using the PARSEC stellar library. Values represent the average [M/H] values from the stars in each age bin where the solutions had a non-zero star formation rate. For time bins that had no reported star formation, no [M/H] value is provided. For Aquarius and Leo~A, we also list systematic uncertainties.}
\end{table*}    

\vspace{0.6in}
\renewcommand\bibname{{References}}
\bibliographystyle{apj}
\bibliography{ms}

\end{document}